\documentclass[11pt]{article}
\usepackage{graphicx}
\usepackage{amsmath}
\title{\bf Production of a gluon with the exchange of three
reggeized gluons in the Lipatov effective action approach.}
\author{M.A.Braun, M.Yu.Salykin, S.S.Pozdnyakov, M.I.Vyazovsky\\
Dep. of High Energy physics,
 Saint-Petersburg State University,\\
198504 S.Petersburg, Russia}

\newcommand\pd{\partial}

\newcommand\ep{\epsilon^*_\perp}

\newcommand\tr{{\rm Tr}}
\newcommand{\Ref}[1]{(\ref{#1})}
\begin{document}

\maketitle

\noindent {\Large\bf Abstract}

In the Regge kinematics the amplitude for gluon production off
three scattering centers is found in the Lipatov effective action
technique. The vertex for gluon emission with the reggeon splitting
in three reggeons is calculated and its transversality is demonstrated.
It is shown that in the sum of all contributions terms
containing principal value singularities are cancelled and substituted by
the standard Feynman poles. These results may be used  for calculation
of the inclusive cross-section for gluon production on two nucleons in the
nucleus.

\section{Introduction}
The inclusive gluon production on the nucleus is one of the
basic processes in high-energy collisions off  nuclear targets.
In the QCD it can be studied either in the Feynman diagram approach
with interacting reggeized gluons ~\cite{braun1} or in the framework
of the dipole
picture (or equivalent  JIMWLK
approach with certain approximations)
in which the gluon field is substituted by  colour dipoles
with their density evolving in rapidity ~\cite{KT}. It is important to compare
these two approaches and see if they are completely equivalent or
different in some respects.
In particular in ~\cite{BSVC} it was advocated that the cross-section
for the gluon production on two centers found in the
framework of the dipole approach in ~\cite{KT} is not complete and
has to be supplemented by new terms involving states composed of three
and four reggeized gluons (the so-called BKP states). On the other
hand in ~\cite{braun2} it was shown that, at least in
lowest orders, contributions from such states in fact cancel, so that
one is left with
exactly the cross-section obtained in the dipole approach.
It should be stressed however that this conclusion was found in the
purely transversal approach with the use of the standard AGK relations
~\cite{AGK} for different cuts of the scattering amplitude off the
two centers. To finally prove the complete equivalence of the
reggeized gluon and dipole approaches one has to compare contributions
from all different cuts and check the validity of the mentioned AGK
rules. This cannot be done in the purely transverse approach used
in ~\cite{braun1,braun2} (nor in the dipole approach) but requires
knowledge of the amplitude for the intermediate gluon production
as a function of its longitudinal momenta. This information is trivial
for production  on  a single scattering center in the target. But it
becomes considerably more complicated when the target involves two or more
centers.

\begin{figure}[h]
\begin{center}
\includegraphics[scale=0.85]{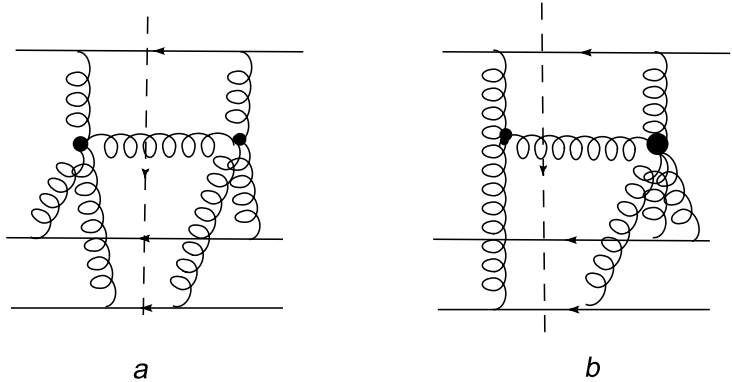}
\end{center}
\caption{Diagrams for the inclusive cross-section on two centers}
\label{fig0}
\end{figure}

A powerful and constructive technique for the calculation of
all Feynman diagrams in the Regge kinematics is provided by
the Lipatov effective action ~\cite{lipatov}.
In the inclusive production on two centers, in the lowest order,
there appear diagrams with production of the observed gluon (P)
from the splitting vertex of a reggeized gluon (R) into two or three
reggeized gluons, shown in
Fig.~\ref{fig0}, $a$ and $b$. The splitting vertex R$\to $RRP
entering Fig.~\ref{fig0}, $a$
was calculated in ~\cite{BV} and its contribution together with all
accompanying ones to the inclusive cross-section was found in
~\cite{BSV}. To finally find the total inclusive cross-section one
has to calculate contributions from diagrams of the type
Fig.~\ref{fig0},$b$ which involve a still more complicated vertex
R$\to$RRRP. This paper is devoted to the study of this vertex and
its contribution together with all accompanying diagrams to the
production amplitude on the right-hand side of the cut
in Fig.~\ref{fig0},$b$.

Our study in ~\cite{BLSV} has found that the vertex for the splitting
of a reggeized gluon into two with emission of a real gluon has
two remarkable properties. The first is that it contains all
contributions in the Regge kinematics for the emitted gluon.
However, contributions from the region in which one of the lower
reggeized gluons has the ''$-$'' component of its momentum much
smaller than the observed gluon lie outside this kinematics and
should be added to the vertex.
The second is that with the latter contribution added
the resulting expression is similar to what is obtained in the
purely transverse picture with addition of standard propagators
for the intermediate quarks and gluons. This observation settles the
problem of light-cone singularities in the Lipatov effective action
formulas and substantially simplifies calculation of the inclusive
cross-sections.

Our results in this paper show that these properties remain valid
when the number of scattering centers is increased  to three.
We suspect that this result is quite general and remains valid for any
number of centers. As for the cross-section to be calculated
from Fig.~\ref{fig0},$b$ we hope that it  will cancel a considerable
part of the calculated terms for
Fig.~\ref{fig0},$a$ and lead to a compact expression for the
total inclusive cross-section. However, this
derivation is postponed for our future publication.

The paper is organized as follows. To make the presentation
self-contained, at least to some extent, in the next section
we reproduce
the main points of the Lipatov effective action formalism and the
known expressions for the R$\to$RP and R$\to$RRP vertices
together with the mentioned properties of the latter, which we are
going to show to be valid also for the R$\to$RRRP vertex.
Section 3  is devoted to the derivation of the
R$\to$RRRP vertex.
In sections 4 and 5 we calculate  diagrams with two and three
interactions of the projectile which are to be added to
the contribution from the R$\to$RRRP vertex.
In section 6 we show that
in the sum of all diagrams one obtains the result which naively
follows from
the purely transverse picture with additional Feynman propagators.
Finally in Section 7 we draw some conclusions. Certain cumbersome
calculations are transferred to Appendices A and B. In Appendix C
we demonstrate that the found R$\to$RRRP vertex is gauge invariant.

In our derivation, as in ~\cite{BV,BLSV,BSV}, for simplicity
we take quarks as the projectile and nuclear components,
invoking colorless exchanges by adequate projectors.

\section{The effective action formalism. R$\to$RP and R$\to$RRP
vertices}

The effective Lagrangian ~\cite{lipatov} describes  interaction
of the quark $\psi$ and  gluon $V_{\mu}=-iT^{a}V^{a}_{\mu}$ fields
and their interaction with the independent reggeon field
$A_{{\mu}}=-iT^{b}A^{b}_{\mu}$ with $A_\perp=0$:
\begin{equation}
{\cal L}_{eff}={\cal L}_{YM}(v_{\mu})
+ \bar{\psi}\Big( i\hat{\pd} + ig\hat{v} -M \Big)\psi
+ \tr\Big(({\cal A}_+(v_+)-A_+)\pd^2_{\perp} A_-\Big)
+ \tr\Big(({\cal A}_-(v_-)-A_-)\pd^2_{\perp} A_+\Big) ,
\label{ei1}
\end{equation}
where ${\cal L}_{YM}$ is the standard Yang-Mills Lagrangian,
$$
{\cal A}_{\pm}(v_{\pm})=-\frac{1}{g}\pd_{\pm}\frac{1}{D_{\pm}}\pd_{\pm}*1=
\sum_{n=0}^{\infty}(-g)^n v_{\pm}(\pd_\pm^{-1}v_\pm)^n
$$
\begin{equation}
=v_{\pm} - gv_{\pm}\pd_\pm^{-1}v_\pm
+ g^2 v_{\pm}\pd_\pm^{-1}v_\pm \pd_\pm^{-1}v_\pm
- g^3 v_{\pm} \pd_\pm^{-1}v_\pm \pd_\pm^{-1}v_\pm \pd_\pm^{-1}v_\pm
+ \dots
\label{ei2}
\end{equation}
and the shift of the field variable $v_\mu =V_\mu +A_\mu$
is done to exclude direct gluon-reggeon transitions.
The unusual two last terms in \Ref{ei1} and correspondent interaction
vertices are called ''induced''.

The Lagrangian is assumed to be local in rapidity, that is, all real
and virtual particles in the direct channels split into groups
in correspondence with their rapidities $y=\frac{1}{2}\ln|p_+/p_-|$
and the Lagrangian \Ref{ei1} describes only interactions
within one group whereas the interaction between groups
with essentially different rapidities
is realized by  reggeon exchange.
The field $A_-$ interacts with a group of a higher rapidity
and the field $A_+$ interacts with a group of a smaller rapidity.
The reggeon propagator (in momentum representation)
\begin{equation}
<A_+^{a}A_-^{b}>=-2i\frac{\delta_{ab}}{q_\perp^2}
\label{ei3}
\end{equation}
corresponds to the exchange of one reggeon.
The following kinematical constraints are implied for the reggeon field
\begin{equation}
\partial_- A_+ =0 , \quad
\partial_+ A_- =0 ,
\label{ei4}
\end{equation}
which reflect the smallness of correspondent components
of reggeon momenta compared to momenta flowing in the group
with the given rapidity.

Note that
the normalization $a_{\pm}=a_0 \pm a_3 = a_{\mu}n^{\pm}_{\mu}$
for longitudinal components of Lorentz vectors is used in this article,
with unit vectors $n^{\pm}_{\mu}=(1,0,0,\mp 1)$, thus
$a_{\mu}b_{\mu}=\frac{1}{2}a_{+}b_{-}+\frac{1}{2}a_{-}b_{+}
+a_{\perp}b_{\perp}$.

Our aim is to calculate the amplitude for the production
of a real gluon with momentum $p$
by the projectile quark of the momentum $k$ with three reggeons
of momenta $q_i$, $i=1,2,3$ interacting with the target quarks.
The multi-Regge kinematical constraints for the case of gluon production
with the momentum $p=q-\sum_i q_{i}$ in the central region are
$$
k_+ =\sqrt{s} \approx k'_+ >> p_+ \approx q_+ >> q_{i+}\approx 0,
$$
$$
\sqrt{s} >> p_- \sim q_{i-} >> q_- = -k'_- \approx 0,
\quad k_- =0,
$$
\begin{equation}
k'_{\perp}=-q_{\perp}
\sim p_{\perp} \sim q_{i\perp} << \sqrt{s} \, ,
\quad  k_{\perp}=0,
\label{ei5}
\end{equation}
where
$k'$ is the momentum of the final quark and $q=k-k'$ is the transferred
momentum.
In this kinematics
the transferred momentum and the reggeon momenta are almost transverse
$q^2 \approx q_{\perp}^2$, $q_{i}^2 \approx q_{i\perp}^2$.
Since the relevant momenta are implied to be very large, we can neglect
all the masses, so the quark is considered massless ($M=0$).

To carry out calculations for the R$\to$RRRP vertex one has to
preliminary know expressions
for the R$\to$RP, and the R$\to$RRP vertices or, in fact, for
the convolution of these vertices with the polarization vector
$\epsilon^{*}$ of the outgoing gluon.
Due to the property of transversality
of reggeon-gluon vertices with respect to the gluon momentum
\cite{lipatov},
the polarization vector obeying the relation $(p\epsilon)=0$
can be chosen independently from the choice of gauge for the gluon field
(we will use the Feynman gauge).
It is most convenient to take the polarization vector satisfying
condition $\epsilon_{+}=0$ and parametrized
by its transverse components:
\begin{equation}
\epsilon_{\mu}(p)=
\epsilon^{\perp}_{\mu}-\frac{(p\epsilon)_{\perp}}{p_{+}} n_{\mu}^{+} ,
\quad \epsilon_{-}(p)=-\frac{2(p\epsilon)_{\perp}}{p_{+}} \ .
\label{eep}
\end{equation}

The R$\to$RP (Lipatov) vertex in a covariant form
is well-known \cite{lipatov}:
\begin{equation}
\frac{g f^{bb_1d}}{2}\left[ q_{\mu}+q_{1\mu}+
\left( \frac{q_1^2}{q_+}-q_{1-} \right) n^+_{\mu}
+\left( \frac{q^2}{q_{1-}}-q_+ \right) n^-_{\mu} \right] ,
\label{elip}
\end{equation}
Its convolution with the polarization vector
takes a simple form
\begin{equation}
-g f^{bb_1d} q_{\perp}^2 L(p,q_1) ,
\label{elipe}
\end{equation}
where
\begin{equation}
L(p,q_1)=\frac{(p\epsilon^{*}_\perp)}{p_\perp^2}-
\frac{(p+q_1,\epsilon^{*}_\perp)}{(p+q_1)_\perp^2} \, .
\label{evl}
\end{equation}
The expression for the R$\to$RRP vertex was derived in \cite{BV}
and its transversality was checked.
Its convolution with the polarization vector can be presented
in the form
\begin{equation}
ig^{2}f^{cb_1 d}f^{db_2 a}
\left[ -\frac{q_{+}B(p,q_{2},q_{1})}{(q-q_{1})^{2}+i0}
+\frac{L(p,q_{2})}{q_{1-}} \right] (p+q_{1}+q_{2})_{\perp}^{2}
+ \Big(\mbox{permutation of } q_{1,2} \mbox{ and } b_{1,2}\Big) ,
\label{ewr}
\end{equation}
where $B$ is the  Bartels vertex
\begin{equation}
B(p,q_{2},q_{1})=L(p+q_2,q_1)=\frac{(p+q_2,\epsilon^{*}_\perp)}{(p+q_2)_\perp^2}-
\frac{(p+q_2+q_1,\epsilon^{*}_\perp)}{(p+q_2+q_1)_\perp^2}\, .
\label{evb}
\end{equation}
The pole in $q_{1-}$ in the second term in square brackets in \Ref{ewr}
is to be understood in the sense of principal value.

\begin{figure}[h]
\begin{center}
\includegraphics[scale=0.85]{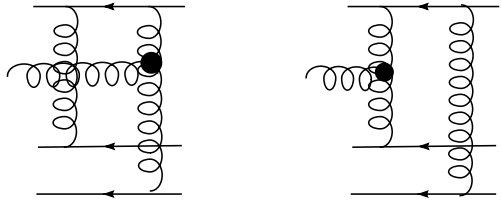}
\end{center}
\caption{Diagrams with a double reggeon exchange.}
\label{fig0a}
\end{figure}

Passing to the production amplitude on the right-hand side of the cut
in Fig. \ref{fig0},$a$ one takes into account that
the vertex for the interaction of a projectile quark
is
\begin{equation}
ig \frac{\gamma_{+}}{2} T^{b} \, .
\label{eq}
\end{equation}
To this production amplitude also two diagrams with a double interaction
of the projectile quark, shown in Fig. \ref{fig0a}, give a contribution,
from which only terms
proportional to $\delta(2k_+q_{1,2-})$ in the projectile quark
propagator should be taken into account \cite{BLSV}.
However, remarkably the rest of the terms in these amplitudes give the
contribution which identically coincides with the second term in
square brackets in \Ref{ewr}. So, as stated in the Introduction, one may
drop this second term containing the principal value singularity and instead
take the diagrams with the double projectile quark interaction
with full Feynman propagator for the intermediate projectile quark.

\section{Effective R$\to$RRRP vertex}

\begin{figure}[h]
\begin{center}
\includegraphics[scale=0.75]{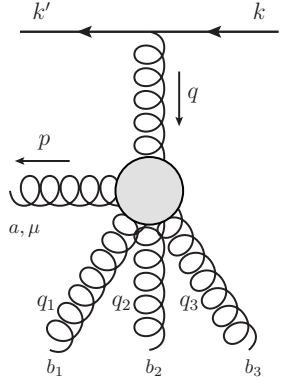}
\end{center}
\caption{A diagram with a single interaction of the projectile.}
\label{fo}
\end{figure}

\begin{figure}[h]
\begin{center}
\includegraphics[scale=0.70]{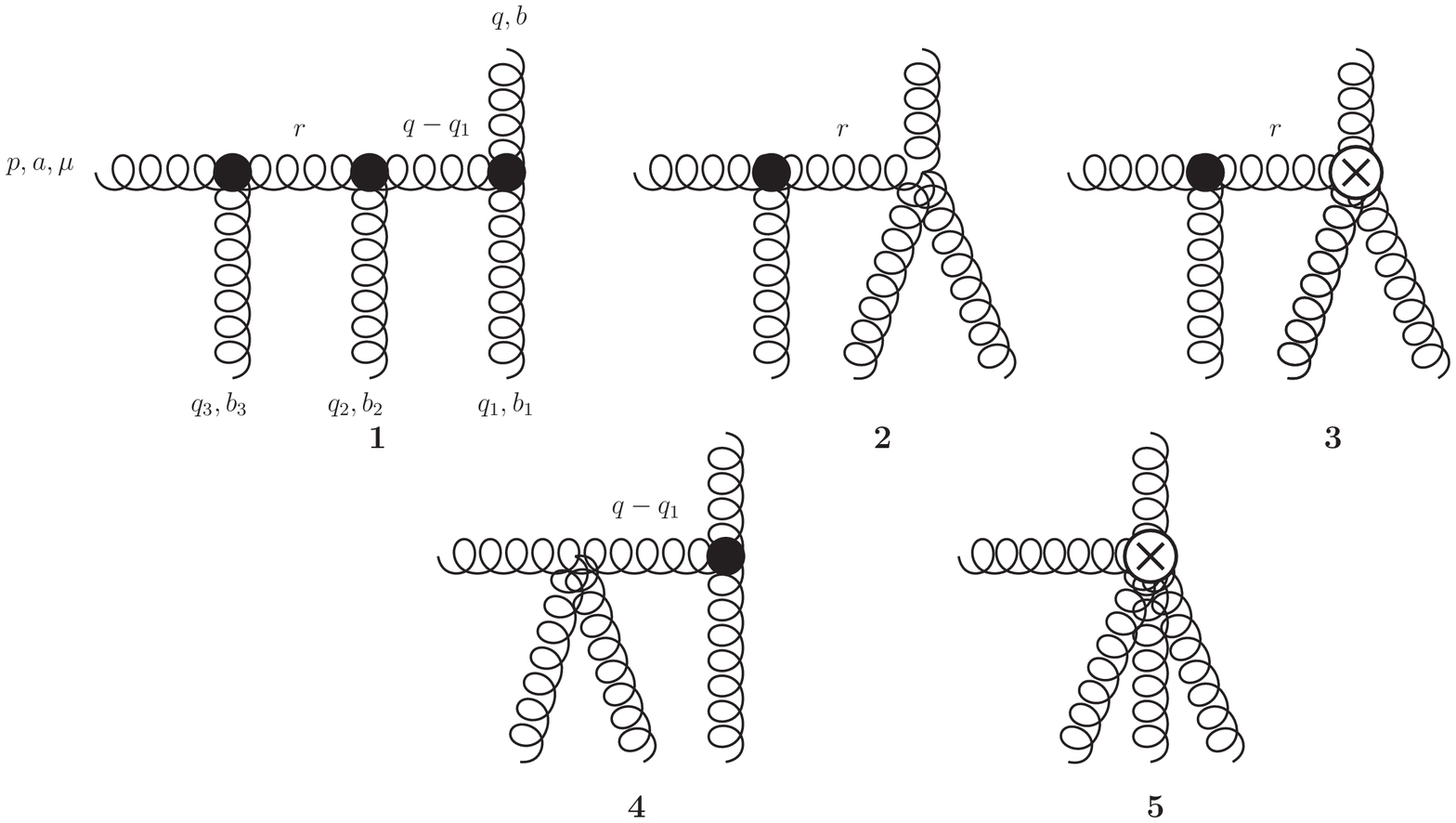}
\end{center}
\caption{Diagrams of the R$\to$RRRP effective vertex.}
\label{fi}
\end{figure}

The contribution of the full effective R$\to$RRRP vertex
to the gluon production amplitude with three reggeons attached to the
target is shown in Fig.~\ref{fo} , where the vertex is marked as
a filled blob. In fact the full vertex is a
sum of several  diagrams of the effective theory which are presented
in Fig.~\ref{fi}.
In all diagrams horizontal curly lines correspond to real or virtual
gluons, vertical curly lines correspond to reggeons and
straight lines with arrows correspond to quarks.
The effective vertex is implied to be symmetrical with respect to all
permutations of the three lower (''induced'') reggeons, so that
full symmetrization
with respect to reggeon momenta and colour indices has to be carried out
for each diagram from Fig.~\ref{fi}.

For brevity in the expressions for all diagrams
the common factor corresponding  to the three propagators
of induced reggeons
$8/(q^2_{1\perp}q^2_{2\perp}q^2_{3\perp})$
will be suppressed in this and subsequent sections.

\subsection{Diagram 1}
Diagram 1 from Fig.~\ref{fi} has to be obtained by convolution
of two ''gluon$\to$reggeon$+$gluon'' (P$\to$RP) vertices
and two propagators of virtual gluons with the Lipatov vertex \Ref{elip}.
For the P$\to$RP vertex we use Eq.(31) from \cite{BV}
which is in our case:
\begin{equation}
\frac{g f^{b_3 ad}}{2} \Big(
-2p_+ g_{\mu\sigma} +(p+2q_3)_{\mu} n^+_{\sigma}
+(p-q_3)_{\sigma} n^+_{\mu} +\frac{q_3^2}{p_+}n^+_{\mu} n^+_{\sigma}
\Big) ,
\label{e11}
\end{equation}
where the virtual gluon momentum $r=q-q_{1}-q_{2}=p+q_3$ has
longitudinal components $r_+=p_+=q_+$ and $r_-=-q_{1-}-q_{2-}$,
and we take gluon propagators in the Feynman gauge.
Thus, the expression for the diagram is found to be
$$
\epsilon^{*}_{\mu} \cdot\frac{g f^{b_3 ad}}{2}
\Big( -2p_+ g_{\mu\sigma} +(p+2q_3)_{\mu} n^+_{\sigma}
+(p-q_3)_{\sigma} n^+_{\mu} +\frac{q_3^2}{p_+}n^+_{\mu}n^+_{\sigma}
\Big)
$$
$$
\times \frac{-i}{(q-q_{1}-q_{2})^2 +i0} \cdot
\frac{g f^{b_2 dc}}{2}
\Big( -2r_+ g_{\sigma\lambda} +(r+2q_2)_{\sigma} n^+_{\lambda}
+(r-q_2)_{\lambda} n^+_{\sigma} +\frac{q_2^2}{r_+}n^+_{\sigma}n^+_{\lambda}
\Big)
$$
$$
\times \frac{-i}{(q-q_{1})^2 +i0} \cdot \frac{g f^{bb_1 c}}{2}
\left[ q_{\lambda}+q_{1\lambda}
+\left( \frac{q_1^2}{q_+} -q_{1-} \right) n^+_{\lambda}
+\left( \frac{q^2}{q_{1-}} -q_+ \right) n^-_{\lambda} \right]
$$
\begin{equation}
=
\frac{g^3 f^{bb_1 c}f^{cb_2 d}f^{db_3 a}}
{((q-q_{1})^2 +i0)((q-q_{1}-q_{2})^2 +i0)}
\left[
-q_+^2 (q_1 \epsilon^* +p\ep +q_2 \epsilon^* +q_3 \epsilon^*)
+ \frac{q_+ q_{\perp}^2}{q_{1-}}\cdot(p+q_2 +q_3)\ep
\right] .
\label{e12}
\end{equation}
Using the identities $(q_i \epsilon)=(q_i \epsilon)_{\perp}$, $i=1,2,3$
and $(p\epsilon)=0$
we get the final expression for the first diagram from Fig.~\ref{fi}:
\begin{equation}
\frac{g^3 f^{bb_1 c}f^{cb_2 d}f^{db_3 a}}
{((q-q_{1})^2 +i0)((q-q_{1}-q_{2})^2 +i0)}
\left[
-q_+^2 (q \epsilon^*)_{\perp}
+ \frac{q_+ q_{\perp}^2}{q_{1-}}\cdot(q-q_1)\ep
\right] ,
\label{e13}
\end{equation}
to which analogous terms with
all simultaneous permutations of momenta $q_{1,2,3}$ and
colour indices $b_{1,2,3}$ of induced reggeons
have to be added.

\subsection{Diagrams 2 and 3}
The diagrams 2 and 3 from Fig.~\ref{fi} are similar, the only difference
between them is that the diagram 2 contains the standard
quadruple vertex
R$\to$RRP  coming from the first term in \Ref{ei1} and diagram 3
contains the induced vertex R$\to$RRP coming from the remaining terms.
To calculate the sum of these diagrams
we need to know the expression for the sum of the quadruple and the
induced vertices, which was already written in \cite{BV}
(Eq.(26) there):
\begin{equation}
\frac{ig^2}{4}\left[
f^{bb_1 c}f^{cb_2 d}
\Big(\frac{2q_\perp^2 n_\sigma^-}{r_-q_{1-}}-n_\sigma^+\Big)
+
f^{bb_2 c}f^{cb_1 d}
\Big(\frac{2q_\perp^2 n_\sigma^-}{r_-q_{2-}}-n_\sigma^+\Big)
\right] ,
\label{e14}
\end{equation}
where $r_-=-q_{1-}-q_{2-}$.
This expression is symmetrical with respect to permutation of reggeons
1 and 2, therefore, taking into account the subsequent
symmetrization over
all permutations of the induced reggeons, it is sufficient to
consider only
the first term in the square brackets (one half of the vertex).

Convoluting the P$\to$RP vertex \Ref{e11}, the gluon propagator
and \Ref{e14}, we get
$$
\epsilon^*_{\mu} \cdot\frac{g f^{b_3 ad}}{2}
\Big( -2p_+ g_{\mu\sigma} +(p+2q_3)_{\mu} n^+_{\sigma}
+(p-q_3)_{\sigma} n^+_{\mu} +\frac{q_3^2}{p_+}n^+_{\mu}n^+_{\sigma}
\Big)
$$
$$
\times
\frac{-i}{(q-q_{1}-q_{2})^2 +i0}
\cdot \frac{ig^2}{4}
f^{bb_1 c}f^{cb_2 d}\left[ -\frac{2q_\perp^2 n_\sigma^-}
{(q_1 +q_2)_{-}q_{1-}} -n_\sigma^+ \right]
$$
\begin{equation}
=-\frac{g^3 f^{bb_1 c}f^{cb_2 d}f^{db_3 a}}
 {(q-q_{1}-q_{2})^2 +i0} \cdot
\frac{q_{\perp}^2 ((q-q_1 -q_2)\ep)}
{q_{1-}(q_{1-} +q_{2-})}\, .
\label{e15}
\end{equation}

\subsection{Diagram 4}
The diagram 4 from Fig.~\ref{fi} contains a new
''gluon$\to$2 reggeons+gluon'' vertex (P$\to$RRP).
The relevant term of the Lagrangian arises from the usual Yang-Mills
quadruple interaction
$$
\frac{g^2}{2} tr\left( [v_{\mu},v_{\lambda}] [v_{\mu},v_{\lambda}] \right)
=\frac{g^2}{4} tr\left( [v_{+},v_{-}] [v_{-},v_{+}] \right) +\cdots
$$
\begin{equation}
=-\frac{g^2}{4} tr\left( [V_{+},A_{-}] [V_{+},A_{-}] \right) +\cdots
=-\frac{g^2}{8}
f^{cb_2 d}f^{db_3 a}
V_{+}^{a} V_{+}^{c} A_{-}^{b_2} A_{-}^{b_3} +\cdots
\label{e16}
\end{equation}
and gives the vertex
\begin{equation}
-i\frac{g^2}{4}
\left( f^{cb_2 d}f^{db_3 a} + f^{cb_3 d}f^{db_2 a} \right)
n^{+}_{\mu}n^{+}_{\lambda}
\label{e17}
\end{equation}
after symmetrization with respect to permutations of gluons and reggeons.
This is the full vertex, since the term $<A_{-}A_{-}V_{\mu}V_{\lambda}>$
is absent in the induced part of \Ref{ei1}.

It turns out that in our gauge the diagram 4 from Fig.~\ref{fi}
gives zero contribution,
because the convolution
$$
\epsilon^{*}_{\mu} \cdot
\frac{-ig^2}{4}
\left( f^{cb_2 d}f^{db_3 a} + f^{cb_3 d}f^{db_2 a} \right)
n^{+}_{\mu}n^{+}_{\lambda}
$$
\begin{equation}
\times \frac{-i}{(q-q_{1})^2 +i0} \cdot \frac{g f^{bb_1 c}}{2}
\left[ q_{\lambda}+q_{1\lambda}
+\left( \frac{q_1^2}{q_+} -q_{1-} \right) n^+_{\lambda}
+\left( \frac{q^2}{q_{1-}} -q_+ \right) n^-_{\lambda} \right]
\label{e18}
\end{equation}
is proportional to $\epsilon^{*}_{+}=0$.

\subsection{Diagram 5}
The diagram 5 from Fig.~\ref{fi} contains the
induced vertex $V_5^{ind}=<A_{+}^{b}(q) V_{-}^{a}(p)
A_{-}^{b_{1}}(q_{1}) A_{-}^{b_{2}}(q_{2}) A_{-}^{b_{3}}(q_{3})>$
(R$\to$RRRP)
arising only from the induced term
\begin{equation}
-g^3 tr(A_{+}\partial^2_{\perp}
v_{-}\partial^{-1}_{-}v_{-}\partial^{-1}_{-}v_{-}\partial^{-1}_{-}v_{-})
\label{a1}
\end{equation}
of the Lagrangian density \Ref{ei1} of the effective theory.
Here $v_{-}=V_{-}+A_{-}$
and the linear in $V_{-}$ part has to be extracted:
$$
-g^3 tr(A_{+}\partial^2_{\perp}
V_{-}\partial^{-1}_{-}A_{-}\partial^{-1}_{-}A_{-}\partial^{-1}_{-}A_{-})
-g^3 tr(A_{+}\partial^2_{\perp}
A_{-}\partial^{-1}_{-}V_{-}\partial^{-1}_{-}A_{-}\partial^{-1}_{-}A_{-})
$$
\begin{equation}
-g^3 tr(A_{+}\partial^2_{\perp}
A_{-}\partial^{-1}_{-}A_{-}\partial^{-1}_{-}V_{-}\partial^{-1}_{-}A_{-})
-g^3 tr(A_{+}\partial^2_{\perp}
A_{-}\partial^{-1}_{-}A_{-}\partial^{-1}_{-}A_{-}\partial^{-1}_{-}V_{-})
\, .
\label{a2}
\end{equation}

It seems to be convenient to denote the momentum $q_4=p$
and the colour index $b_4=a$, then the relation
$q_{1-}+q_{2-}+q_{3-}+q_{4-}=q_{1-}+q_{2-}+q_{3-}+p_{-}=0$
is fulfilled. In this notation the induced vertex is symmetrical
with respect to the simultaneous permutation of corresponding momenta
$q_{1,2,3,4}$ and indices $b_{1,2,3,4}$.
There are 24 terms, the typical one has the form
\begin{equation}
\frac{tr(T^{b}T^{b_1}T^{b_2}T^{b_3}T^{b_4})}
{(q_{2-}+q_{3-}+q_{4-})(q_{3-}+q_{4-})q_{4-}}
=-\frac{tr(T^{b}T^{b_1}T^{b_2}T^{b_3}T^{b_4})}
{q_{1-}(q_{3-}+q_{4-})q_{4-}} \ .
\label{aa3}
\end{equation}
Their calculation is straightforward but requires cumbersome
algebraic transformations of colour factors. So
this calculation is transferred to
Appendix A. The resulting induced vertex is found to be
\begin{equation}
V_5^{ind}=
g^3 \frac{(p\epsilon^{*})_{\perp}}{p^2_{\perp}} q^2_{\perp}
\left(
\frac{f^{bb_1c}f^{cb_2d}f^{db_3a}}{q_{1-}(q_{1-}+q_{2-})}
+ \Big(\mbox{permutations of } q_{1,2,3} \mbox{ and } b_{1,2,3}\Big)
\right) \, .
\label{aafin}
\end{equation}

\subsection{Full effective R$\to$RRRP vertex}
The sum of all diagrams in Fig.~\ref{fi} is thus found to be
$$
g^3 f^{bb_1 c}f^{cb_2 d}f^{db_3 a}
$$
$$
\times \left[
-q_{+}^2 \frac{(q\ep)}{((q-q_{1})^{2}+i0)((q-q_{1}-q_{2})^{2}+i0)}
+q_{+} \frac{q_{\perp}^2 \cdot (q-q_{1})\ep}
  {((q-q_{1})^{2}+i0)((q-q_{1}-q_{2})^{2}+i0) q_{1-}}
\right.
$$
\begin{equation}
\left.
-\frac{q_{\perp}^2 \cdot (q-q_{1}-q_{2})\ep}
  {(q-q_{1}-q_{2})^{2}+i0) (q_{1-}+q_{2-}) q_{1-}}
+\frac{(p\ep)}{p_{\perp}^2}\cdot
 \frac{q_{\perp}^2}{q_{1-} (q_{1-}+q_{2-})}
\right]
\label{e1sum}
\end{equation}
plus 5 other terms with simultaneous permutations of momenta $q_{1,2,3}$
and colour indices $b_{1,2,3}$ of induced reggeons.

One can show that the full R$\to$RRRP vertex is transversal. This is proven
in Appendix C by direct calculation.

For future use we rewrite this result in the manner
similar to the one in \cite{BLSV}, separating terms with causal
and principal value singularities.
Using the identities
$$
\frac{1}{((q-q_{1})^{2}+i0) q_{1-}}=
\frac{q_+}{(q-q_{1})_{\perp}^2 ((q-q_{1})^{2}+i0)}
+\frac{1}{(q-q_{1})_{\perp}^2 q_{1-}} \, ,
$$
\begin{equation}
\frac{1}{((q-q_{1}-q_{2})^{2}+i0) (q_{1-}+q_{2-})}=
\frac{q_+}{(q-q_{1}-q_{2})_{\perp}^2 ((q-q_{1}-q_{2})^{2}+i0)}
+\frac{1}{(q-q_{1}-q_{2})_{\perp}^2 (q_{1-}+q_{2-})} \, ,
\label{e1fr}
\end{equation}
the full contribution from diagrams with the R$\to$RRRP vertex
can be rewritten as a sum of  three terms:
\begin{equation}
A_{I}=g^4  \gamma_{+}
T^{b}f^{bb_1 c}f^{cb_2 d}f^{db_3 a}
\left( W_{I} + Q_{I} + R_{I} \right)
+ \Big(\mbox{permutations of } q_{1,2,3} \mbox{ and } b_{1,2,3}\Big).
\label{e1a}
\end{equation}
Here
\begin{equation}
W_{I}=\frac{q_{+}^{2}B(p,q_{3}+q_{2},q_{1})}
{((q-q_{1})^{2}+i0)((q-q_{1}-q_{2})^{2}+i0)}\, ,
\label{e1w}
\end{equation}
\begin{equation}
Q_{I}=-\frac{q_{+}B(p,q_{3},q_{2})}
{q_{1-}((q-q_{1}-q_{2})^{2}+i0)}\, ,
\label{e1q}
\end{equation}
\begin{equation}
R_{I}=\frac{L(p,q_{3})}{q_{1-}(q_{1-}+q_{2-})}\, .
\label{e1r}
\end{equation}

To finally sum all different contributions in Section 6
it will be convenient to present the colour factor in \Ref{e1a}
in the form
$$
T^{b}f^{bb_1 c}f^{cb_2 d}f^{db_3 a}
=\frac{1}{i}[T^{b_1},T^{c}]f^{cb_2 d}f^{db_3 a}
=-[T^{b_{1}},[T^{b_{2}},T^{d}]]f^{db_{3}a}
$$
\begin{equation}
=(T^{b_2}T^{d}T^{b_1}+T^{b_1}T^{d}T^{b_2}
-T^{d}T^{b_2}T^{b_1}-T^{b_1}T^{b_2}T^{d})f^{db_3 a} .
\label{e1c}
\end{equation}

\subsection{Special kinematical regions}

\begin{figure}[h]
\begin{center}
\includegraphics[scale=0.70]{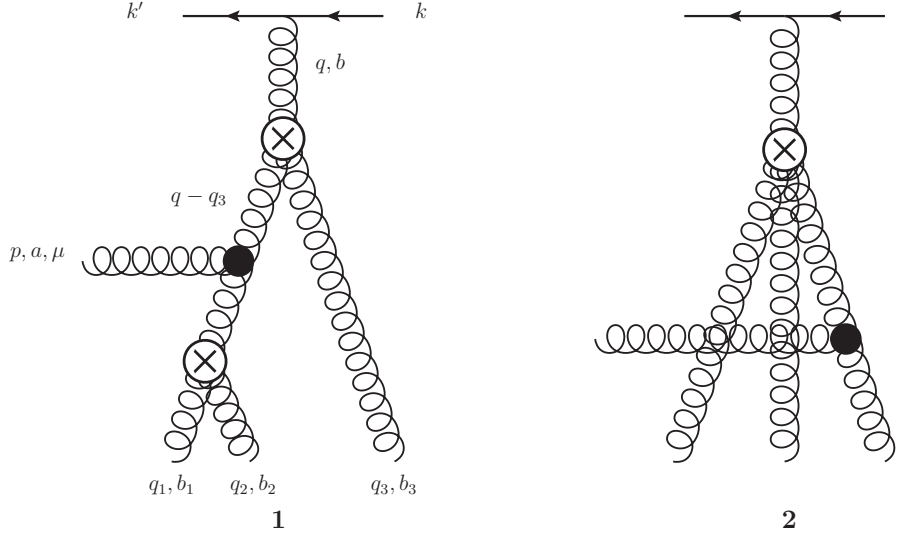}
\end{center}
\caption{Examples of diagrams with the three-reggeon
 and with the four-reggeon vertex.}
\label{multi0}
\end{figure}

\begin{figure}[h]
\begin{center}
\includegraphics[scale=0.70]{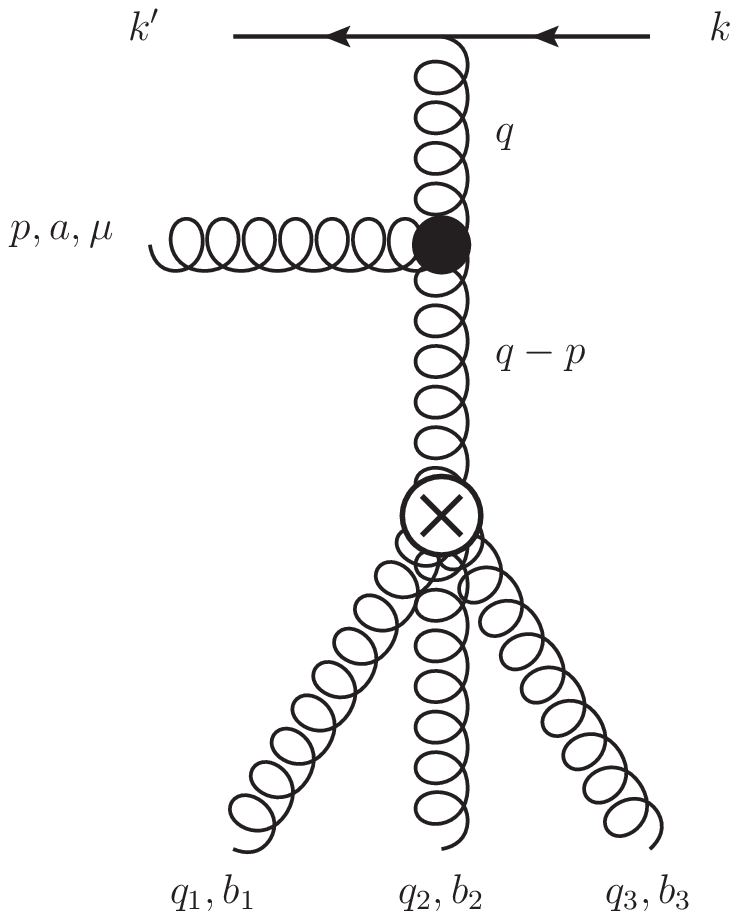}
\end{center}
\caption{A diagram essential in the kinematics $q_{1-}+q_{2-}+q_{3-}=0$.}
\label{multi}
\end{figure}

The vertex R$\to$RRRP calculated in the previous subsection describes
production of a gluon on three centers for arbitrary relations between
$p_-$ and $q_{i-}$, i=1,2,3. In special cases, for particular relations
between these components, it may simplify, degenerating into simpler
diagrams containing triple and quadruple reggeon vertices, as illustrated
in Fig.~\ref{multi0},~\ref{multi}.

For example, under condition
\begin{equation}
|q_{1-}|,|q_{2-}|>>|q_{1-}+q_{2-}| \equiv |q_{3-}+p_-|
\label{e41}
\end{equation}
and as a consequence $q_{1-}+q_{2-}=0$ the general R$\to$RRRP vertex
reduces to two diagrams shown in Fig.~\ref{multi0}.

Of special interest is the kinematical region
in which $p_-<<|q_{i-}|$, $i=1,2,3$.
As argued in \cite{BSV}, this  regime
is essential for the calculation of the inclusive
gluon production from the heavy nucleus.
In this case the momentum $p_{-}$ has to be neglected
compared to all other ''$-$'' components and can be put
to zero.
So one needs the production amplitude in the limit
\begin{equation}
q_{1-}+q_{2-}+q_{3-}=0 ,
\label{e60}
\end{equation}
where $q_{1,2,3-}\neq 0$.

In this limit $q_{+}\approx p_{+}= -\frac{p_{\perp}^2}{p_{-}}$
is very large,
so denominators in Eqs.\Ref{e1w}-\Ref{e1r} can be simplified:
$$
(q-q_1)^2=-q_{+}q_{1-}+(q-q_1)_{\perp}^2 \approx -q_{+}q_{1-} \  ,
$$
\begin{equation}
(q-q_1-q_2)^2=-q_{+}(q_{1-}+q_{2-})+(q-q_1-q_2)_{\perp}^2
\approx -q_{+}(q_{1-}+q_{2-}) .
\label{e61}
\end{equation}
Thus we get
$$
W_{I}+Q_{I}+R_{I}
$$
$$
=
\frac{q_{+}^{2}B(p,q_{3}+q_{2},q_{1})}
{(-q_{+}q_{1-})(-q_{+}(q_{1-}+q_{2-}))}
-\frac{q_{+}B(p,q_{3},q_{2})}
{q_{1-}(-q_{+}(q_{1-}+q_{2-}))}
+\frac{L(p,q_{3})}{q_{1-}(q_{1-}+q_{2-})}
$$
\begin{equation}
=
\frac{B(p,q_{3}+q_{2},q_{1})
+B(p,q_{3},q_{2})+L(p,q_{3})}
{q_{1-}(q_{1-}+q_{2-})} \ ,
\label{e62}
\end{equation}
to which, after multiplying by the colour factor,
all permutations of $q_{1,2,3}$ and $b_{1,2,3}$
have to be added.
Using the definitions of the Lipatov and Bartels vertices
one can write
\begin{equation}
B(p,q_{3}+q_{2},q_{1})+B(p,q_{3},q_{2})+L(p,q_{3})
=\frac{(p\epsilon^{*}_\perp)}{p_\perp^2}-
\frac{(q\epsilon^{*}_\perp)}{q_\perp^2}
=L(p,q_{1}+q_{2}+q_{3}) ,
\label{e63}
\end{equation}
then \Ref{e62} takes the form
\begin{equation}
\frac{L(p,q_{1}+q_{2}+q_{3})}{q_{1-}(q_{1-}+q_{2-})} \ .
\label{e65}
\end{equation}

Taking into account colour factors and the symmetrization
with respect to induced reggeons \Ref{e65} gives
the following contribution to the production amplitude
$$
g^4 \gamma_{+} L(p,q_{1}+q_{2}+q_{3}) T^{b}
\left(
 \frac{f^{bb_1 c}f^{cb_2 d}f^{db_3 a}}{q_{1-}(q_{1-}+q_{2-})}
+\frac{f^{bb_2 c}f^{cb_3 d}f^{db_1 a}}{q_{2-}(q_{2-}+q_{3-})}
\right.
$$
\begin{equation}
\left.
+\frac{f^{bb_3 c}f^{cb_1 d}f^{db_2 a}}{q_{3-}(q_{1-}+q_{3-})}
+\frac{f^{bb_2 c}f^{cb_1 d}f^{db_3 a}}{q_{2-}(q_{1-}+q_{2-})}
+ \frac{f^{bb_1 c}f^{cb_3 d}f^{db_2 a}}{q_{1-}(q_{1-}+q_{3-})}
+\frac{f^{bb_3 c}f^{cb_2 d}f^{db_1 a}}{q_{3-}(q_{2-}+q_{3-})}
\right) \ .
\label{e66}
\end{equation}
If the condition $q_{1-}+q_{2-}+q_{3-}=0$ is fulfilled,
with the use of relations $q_{1-}+q_{3-}=-q_{2-}$,
$q_{2-}+q_{3-}=-q_{1-}$ and
$$
\frac{1}{q_{1-}(q_{1-}+q_{3-})}
=\frac{1}{q_{2-}(q_{2-}+q_{3-})}
= -\frac{1}{q_{1-}(q_{1-}+q_{2-})} -\frac{1}{q_{2-}(q_{1-}+q_{2-})}
$$
one can rewrite \Ref{e66} as the sum of two longitudinal momentum structures:
$$
g^4 \gamma_{+} L(p,q_{1}+q_{2}+q_{3}) T^{b}
\cdot
\frac{f^{bb_1 c}f^{cb_2 d}f^{db_3 a} + f^{bb_3 c}f^{cb_2 d}f^{db_1 a}
- f^{bb_2 c}f^{cb_3 d}f^{db_1 a} - f^{bb_1 c}f^{cb_3 d}f^{db_2 a}}
{q_{1-}(q_{1-}+q_{2-})}
$$
\begin{equation}
+ \Big(\mbox{permutation of } q_{1,2} \mbox{ and } b_{1,2}\Big) .
\label{e67}
\end{equation}
The Jacobi identities
$f^{cb_2 d}f^{db_3 a}=f^{cb_3 d}f^{db_2 a}+f^{cad}f^{db_3 b_2}$
and
$f^{bb_3 c}f^{cb_2 d}=f^{bb_2 c}f^{cb_3 d}-f^{bdc}f^{cb_3 b_2}$
allow us to simplify \Ref{e67}:
$$
g^4 \gamma_{+} L(p,q_{1}+q_{2}+q_{3}) T^{b}
\cdot
\frac{f^{bb_1 c}f^{cad}f^{db_3 b_2} - f^{bdc}f^{cb_3 b_2}f^{db_1 a}}
{q_{1-}(q_{1-}+q_{2-})}
$$
\begin{equation}
+ \Big(\mbox{permutation of } q_{1,2} \mbox{ and } b_{1,2}\Big) .
\label{e68}
\end{equation}
If we interchange indices of summation $c$ and $d$
in the second term of the numerator, we get the expression
proportional to another Jacobi identity
$f^{bb_1 c}f^{cad}-f^{bcd}f^{cb_1 a}=f^{bac}f^{cb_1 d}$,
and then the full contribution to the amplitude from diagrams
with the R$\to$RRRP vertex in the kinematics $q_{1-}+q_{2-}+q_{3-}=0$
takes the form
\begin{equation}
g^4 \gamma_{+} L(p,q_{1}+q_{2}+q_{3}) T^{b}
\frac{f^{bac}f^{cb_1 d}f^{db_3 b_2}}{q_{1-}(q_{1-}+q_{2-})}
+ \Big(\mbox{permutation of } q_{1,2} \mbox{ and } b_{1,2}\Big) .
\label{e6a}
\end{equation}
This expression exactly coincides with the contribution
of the diagram from Fig.~\ref{multi} calculated in Appendix B.

In general, diagrams with multi-reggeon effective vertices
give a contribution only when the reggeon momenta are
restricted by specific kinematical conditions (see \cite{BLSV}).
To find these conditions, the equations
$\partial_{-}A_{+}=0$ and $\partial_{+} A_{-}=0$
have to be applied to reggeon fields
interacting by means of the multi-reggeon vertex.
For example, for the first diagram from Fig.~\ref{multi0}
with the use of the equation $\partial_{-} A_{+}=0$
for the three-reggeon vertex we get the momentum condition
\begin{equation}
q_{1-}+q_{2-}=0 ,
\label{e41x}
\end{equation}
which should be understood not as strict equation
but as the restriction \Ref{e41}.
From the equation $\partial_{+} A_{-} =0$
the conditions $q_{1+}=0$ and $q_{2+}=0$ follow
which are always assumed in the Regge kinematics.
The relevant momentum condition for the diagram from Fig.~\ref{multi}
with the vertex R$\to$RRR is \Ref{e60} and it is the only diagram
for this kinematics.

\section{Double interaction of the projectile}

\begin{figure}[h]
\begin{center}
\includegraphics[scale=0.75]{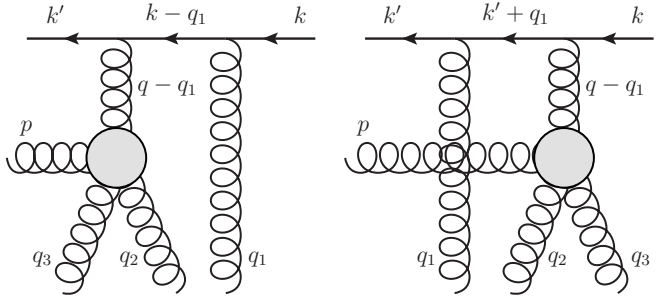}
\end{center}
\caption{Diagrams with a double interaction of the projectile
and the gluon emitted from the R$\to$RRP vertex.}
\label{fii}
\end{figure}
\begin{figure}[h]
\begin{center}
\includegraphics[scale=0.75]{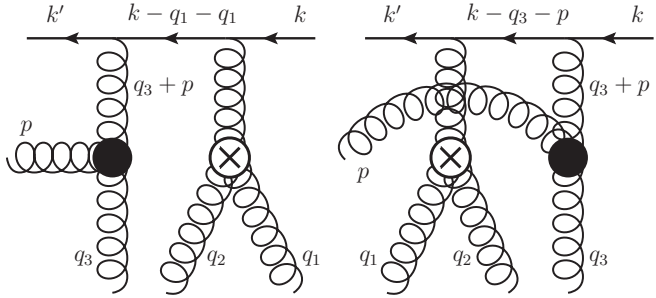}
\end{center}
\caption{Diagrams with a double interactions of the projectile and
the gluon emitted from the R$\to$RP vertex.}
\label{fiv}
\end{figure}

Diagrams with two interactions of the projectile separate into two
groups with emission of the gluon from the R$\to$RRP vertex
shown in Fig. \ref{fii} and from a reggeon shown in Fig. \ref{fiv}.

\subsection{Diagrams in Fig. \ref{fii}}

We start with diagrams in Fig.~\ref{fii}.
In both diagrams the filled blob corresponds to the R$\to$RRP vertex.
Since the gluon is produced by this vertex we need to know only
the convolution of the vertex with the polarization vector.
It is given by \Ref{ewr},
for diagrams from Fig.~\ref{fii} the vertex is
\begin{equation}
ig^{2}f^{cb_{2}d}f^{db_{3}a}
\left[
-\frac{q_{+}B(p,q_{3},q_{2})}{(q-q_{1}-q_{2})^{2}+i0}
+\frac{L(p,q_{3})}{q_{2-}}
\right]
 (q-q_{1})_{\perp}^{2}
+ \Big(\mbox{permutation of } q_{2,3} \mbox{ and } b_{2,3}\Big)\, .
\label{e2wr}
\end{equation}
Since the R$\to$RRP vertex is already symmetrical with respect to
permutation of the two induced reggeons, the symmetrization for
diagrams in Fig.~\ref{fii} have to be carried out only
for cyclic permutations of momenta $q_{1,2,3}$ and
colour indices $b_{1,2,3}$.

Note an important point.
As argued in \cite{BLSV}, from the property of  locality in rapidity,
it follows that propagators of the virtual quark in diagrams such as
in Fig.~\ref{fii} are to  be substituted by
their parts proportional to the $\delta$-function.
Thus, we have to write for the quark propagators
$$
\frac{k_{+}}{(k-q_{1})^{2}+i0}
\approx \frac{k_{+}}{-k_{+}q_{1-}+(k-q_{1})_{\perp}^{2}+i0}
$$
\begin{equation}
\approx \frac{k_{+}}{-k_{+}q_{1-}+i0}
=-P\frac{1}{q_{1-}}-i \pi \delta (q_{1-})
\to -i\pi\delta(q_{1-})
\label{ep1}
\end{equation}
and
$$
\frac{k'_{+}}{(k'+q_{1})^{2}+i0}
\approx \frac{k_{+}}{k_{+}q_{1-}+(k'+q_1)_{\perp}^{2}+i0}
$$
\begin{equation}
\approx \frac{k_{+}}{k_{+}q_{1-}+i0}
=P\frac{1}{q_{1-}}-i \pi \delta (q_{1-})
\to -i\pi\delta(q_{1-}) .
\label{ep2}
\end{equation}
The reason is that the application of the condition of the similarity
of rapidities of the scattered quarks and the virtual quark is
equivalent to the momentum cut-off which makes the pole term
of the propagator tend to zero in the sense of generalized function.
The remaining term with the delta-function $-i\pi\delta (q_{1-})$
can be treated as the effective propagator
of a quark moving with a high rapidity.

The fermionic part of the first diagram from Fig.~\ref{fii} including
the numerator of quark propagators and quark-reggeon vertices is
\begin{equation}
(ig)^{2} \frac{\gamma_{+}}{2}
i(\hat{k}-\hat{q_{1}})\frac{\gamma_{+}}{2}
=-ig^{2}\frac{\gamma_{+}}{2}(k_{+}-q_{1+})
\approx \frac{-i}{2} g^2 k_{+}\gamma_{+}\, ,
\label{e20}
\end{equation}
where the identities $\gamma_{+}\gamma_{+}=0$,
$\gamma_{+}\gamma_{-}\gamma_{+}=4\gamma_{+}$ are used.
The fermionic part of the second diagram from Fig.~\ref{fii} is
identical to \Ref{e20}.
Multiplying this expression by the propagator of virtual reggeon
$-2i/(q-q_{1})_{\perp}^{2}$
and by the common factor $(q-q_{1})_{\perp}^{2}$ from \Ref{e2wr}
we get $-g^2 k_{+}\gamma_{+}$.

Since the effective quark propagators are the same for both diagrams
in Fig.~\ref{fii}, their momentum factors are
identical and these diagrams differ only by colour factors,
which are
\begin{equation}
T^{c}T^{b_1}f^{cb_2 d}f^{db_3 a}
=-i [T^{b_2},T^{d}] T^{b_1} f^{db_3 a}
=-i f^{db_3 a} (T^{b_2}T^{d}T^{b_1}-T^{d}T^{b_2}T^{b_1}) .
\label{e23}
\end{equation}
for the first diagram and
\begin{equation}
T^{b_1}T^{c}f^{cb_2 d}f^{db_3 a}
=-i T^{b_1}[T^{b_2},T^{d}] f^{db_3 a}
=-i f^{db_3 a} (T^{b_1}T^{b_2}T^{d}-T^{b_1}T^{d}T^{b_2})
\label{e24}
\end{equation}
for the second one.

Collecting all factors and taking into account the symmetrization
with respect to induced reggeons, we get the full result for
the contribution from diagrams with a double interaction
of the projectile and the gluon emitted from the R$\to$RRP vertex:
$$
g^{4}  \gamma_{+}
f^{db_{3}a}  (T^{b_2}T^{d}T^{b_1}-T^{d}T^{b_2}T^{b_1}
+T^{b_1}T^{b_2}T^{d}-T^{b_1}T^{d}T^{b_2})
$$
$$
\times
\left[ \frac{q_{+}B(p,q_{3},q_{2})}{(q-q_{1}-q_{2})^{2}+i0}
-\frac{L(p,q_{3})}{q_{2-}} \right]
 ( -i\pi\delta(q_{1-}) )
$$
\begin{equation}
+ \Big(\mbox{permutations of } q_{1,2,3} \mbox{ and } b_{1,2,3}\Big) .
\label{e2a}
\end{equation}
In the following the first term in square brackets in \Ref{e2a}
is referred to as $W_{II}$ and the second one as $R_{II}$.

\subsection{Diagrams in Fig. \ref{fiv}}

Passing to the diagrams in Fig. \ref{fiv} (with all permutations
of the induced reggeons) we note that
the three-reggeon vertex R$\to$RR was calculated in \cite{BLSV},
but the other normalization of the ''$\pm$'' components
was chosen there, so we re-derive the expression.
The relevant term of the effective Lagrangian which arises
only from the induced term is
$$(-g) tr(A_{+} \pd_{\perp}^2 A_{-} \pd_{-}^{-1} A_{-} )
=-\frac{g}{2} tr(A_{+} \pd_{\perp}^2 [A_{-},(\pd_{-}^{-1}A_{-})])
$$
\begin{equation}
=\frac{g}{4}f^{cb_1 b_2}
A_{+}^{c} \pd_{\perp}^{2}
A_{-}^{b_1}(\pd_{-}^{-1} A_{-}^{b_2})\, .
\label{e44}
\end{equation}
The vertex in momentum representation
\begin{equation}
\frac{gf^{cb_1 b_2}}{4}\, (q_1 + q_2)_{\perp}^{2}
\left(\frac{1}{q_{1-}}-\frac{1}{q_{2-}}\right)
\label{e45}
\end{equation}
is symmetrical with respect to the permutation of the reggeons 1 and 2.
Using the condition \Ref{e41x}, we can rewrite the vertex \Ref{e45}
in the form
\begin{equation}
\frac{gf^{cb_1 b_2}}{2q_{1-}} (q_1 + q_2)_{\perp}^{2}
=-\frac{gf^{cb_1 b_2}}{2q_{2-}} (q_1 + q_2)_{\perp}^{2} \, .
\label{e46}
\end{equation}
As stated in \cite{BLSV}, the poles in $q_{1-}$ and $q_{2-}$
should be understood in the sense of principal value.
Although the three-reggeon vertex gives a non-zero contribution
only under condition $q_{1-}+q_{2-}=0$, in the presence of the factor
$ -i\pi \delta(q_{1-}+q_{2-})$ remaining from the propagator
of the virtual quark this condition is always fulfilled.
Therefore one have to include the contribution of diagrams
from Fig.~\ref{fiv} in a general Regge kinematics.

The fermionic part of both diagrams from Fig.~\ref{fiv} is the same
as for diagrams of Section 4.1:
\begin{equation}
\frac{-i}{2} g^{2} k_{+} \gamma_{+}
\label{e47}
\end{equation}
and the quark propagators are substituted by their delta-functional
parts:
$$
\frac{k_{+}}{(k-q_{1}-q_{2})^{2}+i0} \approx
\frac{1}{-(q_{1-}+q_{2-})+i0} \to -i\pi \delta(q_{1-}+q_{2-})
\mbox{ and}
$$
\begin{equation}
\frac{k'_{+}}{(k-q_{3}-p)^{2}+i0}
=\frac{k'_{+}}{(k'+q_{1}+q_{2})^{2}+i0} \approx
\frac{1}{(q_{1-}+q_{2-})+i0} \to -i\pi \delta(q_{1}+q_{2})
\label{e48}
\end{equation}
for the first and the second diagram respectively.
Multiplying this by two propagators of virtual reggeons,
the Lipatov vertex
\begin{equation}
-g f^{db_{3}a} (p+q_{3})^{2}_{\perp} L(p,q_3)
\label{e31a}
\end{equation}
and by the three-reggeon vertex \Ref{e46}
we get the answer
$$
-ig^{4} \gamma_{+}
f^{db_{3}a}f^{cb_{1}b_{2}} \left( T^{d}T^{c}+T^{c}T^{d} \right)
\frac { L(p,q_{3})}{q_{1-}}
( -i\pi\delta(q_{1-}+q_{2-}) )
$$
\begin{equation}
=g^{4}  \gamma_{+} f^{db_{3}a}
\left( T^{d}T^{b_2}T^{b_1}-T^{d}T^{b_1}T^{b_2}
+T^{b_2}T^{b_1}T^{d}-T^{b_1}T^{b_2}T^{d} \right)
\frac{L(p,q_{3})}{q_{1-}}
( -i\pi\delta(q_{1-}+q_{2-}) ) ,
\label{e49}
\end{equation}
which is symmetrical with respect to permutation of the reggeons 1 and 2.
Under the symmetrization over all permutations of the induced
reggeons the full contribution from diagrams with the three-reggeon
vertex can be written in the form
$$
A_{IV}=g^{4}  \gamma_{+} f^{db_{3}a}
\left( \frac{T^{d}T^{b_2}T^{b_1}}{q_{1-}}
- \frac{T^{b_2}T^{b_1}T^{d}}{q_{2-}} \right)
L(p,q_{3}) (-i \pi\delta(q_{1-}+q_{2-}))
$$
\begin{equation}
+ \Big(\mbox{permutations of } 1,2,3\Big) .
\label{e4a}
\end{equation}

\section{Triple interaction of the projectile}

\begin{figure}[h]
\begin{center}
\includegraphics[scale=0.70]{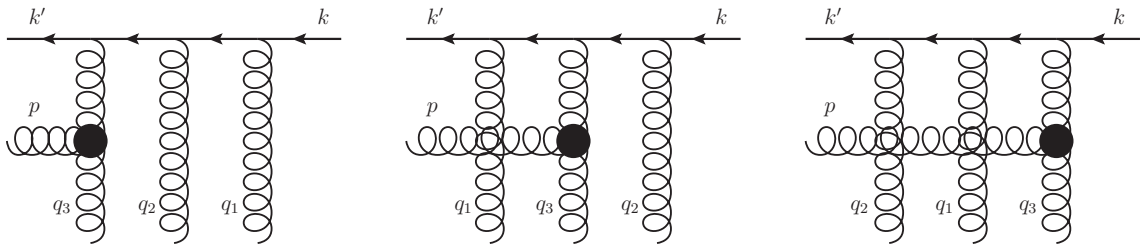}
\end{center}
\caption{Diagrams with a triple interaction of the projectile.}
\label{fiii}
\end{figure}

There are 18 diagrams with three reggeons interacting with the projectile
quark. Three of them are shown in Fig.~\ref{fiii}, the others can be
obtained by means of all permutations of reggeons 1,2,3.
The fermionic part of the first diagram from Fig.~\ref{fiii} including
the numerators of quark propagators and quark-reggeon vertices is
$$
(ig)^{3}
\frac{\gamma_{+}}{2}i(\hat{k}-\hat{q_{1}}-\hat{q_{2}})
\frac{\gamma_{+}}{2}i(\hat{k}-\hat{q_{1}})\frac{\gamma_{+}}{2}
$$
\begin{equation}
=ig^{3}
\frac{\gamma_{+}}{2}
\left( \frac{1}{2}(k_{+}-q_{1+}-q_{2+})\gamma_{-} \right)
\frac{\gamma_{+}}{2}
\left( \frac{1}{2}(k_{+}-q_{1+})\gamma_{-} \right)
\frac{\gamma_{+}}{2}
=\frac{i}{2} g^{3} k_{+}^{2} \gamma_{+}\, .
\label{e31}
\end{equation}
The other diagrams in Fig.~\ref{fiii} give the same result
in the approximation of the Regge kinematics.
It has to be multiplied by the propagator of virtual reggeon
${-2i}/{(p+q_{3})^{2}_{\perp}}$
and by the Lipatov vertex \Ref{e31a}.

The condition of locality in rapidity
has to be applied to the two virtual quarks independently.
In correspondence with our notation in the previous section,
the full propagator of each quark has to be
substituted by its delta-functional part:
$$
\frac{k_{+}^{2}}{((k-q_{1}-q_{2})^{2}+i0)((k-q_{1})^{2}+i0)}
\approx \frac{1}{(-(q_{1-}+q_{2-})+i0)(-q_{1-}+i0)}
\to (-i\pi)^2 \delta(q_{1-}+q_{2-})\delta(q_{1-}) ,
$$
$$
\frac{k^{2}_{+}}{((k'+q_{1})^{2}+i0)((k-q_{2})^{2}+i0)}
\approx \frac{1}{(q_{1-}+i0)(-q_{2-}+i0)}
\to (-i\pi)^2 \delta(q_{1-})\delta(q_{2-}) ,
$$
\begin{equation}
\frac{k^{2}_{+}}{((k'+q_{2})^{2}+i0)((k'+q_{1}+q_{2})^{2}+i0)}
\approx \frac{1}{(q_{2-}+i0)((q_{1-}+q_{2-})+i0)}
\to (-i\pi)^2 \delta(q_{2-})\delta(q_{1-}+q_{2-})
\label{e32}
\end{equation}
for the first, second and third diagram from Fig.~\ref{fiii} respectively.
Since
\begin{equation}
\delta(q_{1-})\delta(q_{1-}+q_{2-})=\delta(q_{2-})\delta(q_{1-}+q_{2-})
=\delta(q_{1-})\delta(q_{2-})\, ,
\label{e33}
\end{equation}
one can see that the momentum factors are identical for the three
diagrams in Fig.~\ref{fiii}.

Taking into account the colour factors of diagrams
in Fig.~\ref{fiii} and the symmetry with respect to all
permutations of induced reggeons, we get the full contribution
from diagrams with the three reggeon exchange:
$$
A_{III}=-g^{4}\gamma_{+}
f^{db_{3}a} (T^{d}T^{b_2}T^{b_1}+T^{b_1}T^{d}T^{b_2}+T^{b_2}T^{b_1}T^{d})
L(p,q_{3}) (-i\pi)^2 \delta(q_{1-})\delta(q_{2-})
$$
\begin{equation}
+ \Big(\mbox{permutations of } 1,2,3\Big) .
\label{e34}
\end{equation}

\section{Restoration of causal propagators}

Separate parts of the calculated production amplitude contain
poles  in the ''$-$'' longitudinal momenta or their sums, coming from
the similar singularity of the induced part of the effective action,
whose exact meaning is not fixed {\it apriori}. There was a special
discussion on this point in ~\cite{hentschinski1, hentschinski2},
which however was restricted to the simplest case with only one center
in the target. In ~\cite{BLSV} for the case of two centers in the
target, it was established that the correct
reproduction of the relevant Feynman diagrams requires
these singularities to be interpreted in the principal value sense.
Moreover, it was shown that these singularities combined with the
$\delta$-functional contributions from  double interactions of the
projectile turn into the latter with normal causal propagators.
In this section, assuming that singularities in ''$-$'' components
of the momenta are to be interpreted in the principal value sense,
we demonstrate that with three centers in the target they also
combine into the final expression containing only usual singularities
of Feynman diagrams, changing the $\delta$-functions in the
propagators of the projectile quark into normal causal propagators.

From diagrams with the single interaction of the projectile
we get contribution  $W_{I}$ \Ref{e1w}
\begin{equation}
g^{4} \gamma_{+} T^{b}f^{bb_{1}c}f^{cb_{2}d}f^{db_{3}a}
\frac{q_{+}^{2}\, B(p,q_{3}+q_{2},q_{1})}
{((q-q_{1})^{2}+i0)((q-q_{1}-q_{2})^{2}+i0)}
+ \Big(\mbox{permutations of } 1,2,3\Big) ,
\label{e51}
\end{equation}
which contains only usual Feynman poles.

Next, the principal value pole from the part $Q_{I}$ \Ref{e1q}
and the delta-function contribution $W_{II}$ (the first term
in square brackets in \Ref{e2a}) from diagrams
with a double interaction of the projectile combine into the
Feynman pole of the causal propagator:
$$
g^{4}\gamma_{+}f^{db_3 a}
\frac{q_{+}B(p,q_{3},q_{2})}{(q-q_{1}-q_{2})^{2}+i0}\,
\left[ \left(T^{b_2}T^{d}T^{b_1}-T^{d}T^{b_2}T^{b_1} \right)
\left( -P\frac{1}{q_{1-}} -i\pi\delta(q_{1-}) \right) \right.
$$
$$
+\left. \left( T^{b_1}T^{b_2}T^{d}-T^{b_1}T^{d}T^{b_2} \right)
\left( P\frac{1}{q_{1-}} -i\pi\delta(q_{1-}) \right) \right]
+ \Big(\mbox{permutations of } 1,2,3\Big)
$$
\begin{equation}
\approx g^{4}  \gamma_{+} f^{db_{3}a}
\frac{q_{+}B(p,q_{3},q_{2})}{(q-q_{1}-q_{2-})^{2}+i0}\!
\left[ \frac{k_{+}\cdot if^{cb_2 d}T^{c}T^{b_1}}{(k-q_{1})^{2}+i0}
+\frac{k_{+}\cdot if^{cb_2 d}T^{b_1}T^{c}}{(k'+q_{1})^{2}+i0}
\right]
\!+\! \Big(\mbox{permutations of } 1,2,3\Big) .
\label{e52}
\end{equation}
This expression corresponds to the sum of diagrams in Fig.~\ref{fii}
with only the part $W$ of the effective vertex R$\to$RRP taken into
account and with the quark propagators taken as a whole.

Finally, we check that the remaining terms,
that is the contribution from $R_{I}+R_{II}+A_{III}+A_{IV}$,
combine into the expression containing only causal propagators.
Note that each of these terms is proportional to the Lipatov vertex.
Because of the symmetry with respect to permutation of the induced
reggeons, it is sufficient to consider only  terms proportional
to $L(p,q_3)$. In subsequent equations all poles in $q_{1,2-}$
and $q_{1-}+q_{2-}$ are understood in the sense of principal value.
Then, suppressing the common factor
$g^{4}\gamma_{+}f^{db_3 a}L(p,q_3)$ we have:
$$
R_{I}=
\frac{T^{b_2}T^{d}T^{b_1}+T^{b_1}T^{d}T^{b_2}
-T^{d}T^{b_2}T^{b_1}-T^{b_1}T^{b_2}T^{d}} {q_{1-}(q_{1-}+q_{2-})}
+ \Big(\mbox{permutation of } 1,2\Big) ,
$$
$$
R_{II}=
-\frac{T^{b_2}T^{d}T^{b_1}-T^{d}T^{b_2}T^{b_1}
+T^{b_1}T^{b_2}T^{d}-T^{b_1}T^{d}T^{b_2}} {q_{2-}}
(-i\pi)\delta(q_{1-})
+ \Big(\mbox{permutation of } 1,2\Big) ,
$$
$$
A_{III}=
-(T^{d}T^{b_2}T^{b_1}+T^{b_1}T^{d}T^{b_2}+T^{b_2}T^{b_1}T^{d})
(-i\pi)^2 \delta(q_{1-})\delta(q_{2-})
+ \Big(\mbox{permutation of } 1,2\Big) ,
$$
\begin{equation}
A_{IV}=
\left( \frac{T^{d}T^{b_2}T^{b_1}}{q_{1-}}
-\frac{T^{b_2}T^{b_1}T^{d}}{q_{2-}} \right)
(-i\pi)\delta(q_{1-}+q_{2-})
+ \Big(\mbox{permutation of } 1,2\Big) .
\label{e53}
\end{equation}
The sum is
$$\frac{T^{b_2}T^{d}T^{b_1}+T^{b_1}T^{d}T^{b_2}
  -T^{d}T^{b_2}T^{b_1}-T^{b_1}T^{b_2}T^{d}} {q_{1-}(q_{1-}+q_{2-})}
+ \frac{T^{b_1}T^{d}T^{b_2}+T^{b_2}T^{d}T^{b_1}
  -T^{d}T^{b_1}T^{b_2}-T^{b_2}T^{b_1}T^{d}} {q_{2-}(q_{1-}+q_{2-})}
$$
$$
+\frac{-T^{b_2}T^{d}T^{b_1}+T^{d}T^{b_2}T^{b_1}
  -T^{b_1}T^{b_2}T^{d}+T^{b_1}T^{d}T^{b_2}} {q_{2-}}
(-i\pi)\delta(q_{1-})
$$
$$
+\frac{-T^{b_1}T^{d}T^{b_2}+T^{d}T^{b_1}T^{b_2}
  -T^{b_2}T^{b_1}T^{d}+T^{b_2}T^{d}T^{b_1}} {q_{1-}}
(-i\pi)\delta(q_{2-})
$$
$$
+ \frac{T^{d}T^{b_2}T^{b_1}-T^{b_1}T^{b_2}T^{d}} {q_{1-}}
(-i\pi)\delta(q_{1-}+q_{2-})
+ \frac{T^{d}T^{b_1}T^{b_2} -T^{b_2}T^{b_1}T^{d}} {q_{2-}}
(-i\pi)\delta(q_{1-}+q_{2-})
$$
\begin{equation}
-(T^{d}T^{b_2}T^{b_1}+T^{b_1}T^{d}T^{b_2}
+T^{b_2}T^{b_1}T^{d}+T^{d}T^{b_1}T^{b_2}
+T^{b_2}T^{d}T^{b_1}+T^{b_1}T^{b_2}T^{d})
 (-i\pi)^2 \delta(q_{1-})\delta(q_{2-}) .
\label{e54}
\end{equation}
With the use of the following identities in the sense
of the generalized functions
$$
\delta(q_{1-}) P\frac{1}{q_{2-}} =\delta(q_{1-}) P\frac{1}{q_{1-}+q_{2-}}\, ,
\quad
\delta(q_{2-}) P\frac{1}{q_{1-}} =\delta(q_{2-}) P\frac{1}{q_{1-}+q_{2-}}\, ,
$$
$$
\delta(q_{1-}+q_{2-}) P\frac{1}{q_{2-}}
=-\delta(q_{1-}+q_{2-}) P\frac{1}{q_{1-}}\, ,
$$
$$
\delta(q_{1-}) \delta(q_{2-})
=\delta(q_{1-}) \delta(q_{2-}+q_{1-})
=\delta(q_{1-}+q_{2-}) \delta(q_{2-}) ,
$$
$$
P\frac{1}{q_{1-}}\cdot P\frac{1}{q_{1-}+q_{2-}}
+ P\frac{1}{q_{2-}}\cdot P\frac{1}{q_{1-}+q_{2-}}
= P\frac{1}{q_{1-}}\cdot P\frac{1}{q_{2-}}
$$
one can write \Ref{e54}, separating different colour
structures, as
$$
T^{d}T^{b_2}T^{b_1}
\left(-P\frac{1}{q_{1-}}\cdot P\frac{1}{q_{1-}+q_{2-}}
+ (-i\pi)\delta(q_{1-}) P\frac{1}{q_{1-}+q_{2-}}
\right. $$ $$
\quad\quad \quad\quad \left.
+ (-i\pi)\delta(q_{1-}+q_{2-}) P\frac{1}{q_{1-}}
- (-i\pi)^2 \delta(q_{1-}+q_{2-})\delta(q_{1-}) \right)
$$
$$
+ T^{b_1}T^{d}T^{b_2}
\left( P\frac{1}{q_{1-}}\cdot P\frac{1}{q_{2-}}
+ (-i\pi)\delta(q_{1-}) P\frac{1}{q_{2-}}
- (-i\pi)\delta(q_{2-}) P\frac{1}{q_{1-}}
- (-i\pi)^2 \delta(q_{1-})\delta(q_{2-}) \right)
$$
$$
+ T^{b_2}T^{b_1}T^{d}
\left( -P\frac{1}{q_{2-}}\cdot P\frac{1}{q_{1-}+q_{2-}}
- (-i\pi)\delta(q_{2-}) P\frac{1}{q_{1-}+q_{2-}}
\right.
$$
\begin{equation}
\quad\quad \quad\quad \left.
- (-i\pi)\delta(q_{1-}+q_{2-}) P\frac{1}{q_{2-}}
- (-i\pi)^2 \delta(q_{1-}+q_{2-})\delta(q_{2-}) \right)
+ \Big(\mbox{permutation of } 1,2 \Big)
\label{e55}
\end{equation}
$$
= -\frac{T^{d}T^{b_2}T^{b_1}}
 { (-(q_{1-}+q_{2-})+i0) (-q_{1-}+i0) }
-\frac{T^{b_1}T^{d}T^{b_2}}
 { (q_{1-}+i0) (-q_{2-}+i0)}
$$
\begin{equation}
-\frac{T^{b_2}T^{b_1}T^{d}}
 { (q_{2-}+i0) ((q_{1-}+q_{2-})+i0) }
+ \Big(\mbox{permutation of } 1,2\Big) .
\label{e56}
\end{equation}

In the approximation of the Regge kinematics
the full contribution from $R_{I}+R_{II}+A_{III}+A_{IV}$
is
$$
-g^{4}\gamma_{+}f^{db_3 a}
\left[ \frac{k_{+}^{2}\cdot T^{d}T^{b_2}T^{b_1}}
 { ((k-q_{1}-q_{2})^{2}+i0) ((k-q_{1})^{2}+i0) }
+ \frac{k_{+}^{2}\cdot T^{b_1}T^{d}T^{b_2}}
 { ((k'+q_{1})^{2}+i0) ((k-q_{2})^{2}+i0)}
\right.
$$
\begin{equation}
+ \left. \frac{k_{+}^{2}\cdot T^{b_{2}}T^{b_{1}}T^{d}}
 { ((k'+q_{2})^{2}+i0) ((k'+q_{1}+q_{2})^{2}+i0) } \right]
L(p,q_3)
+ \Big(\mbox{permutations of } 1,2,3\Big) .
\label{e57}
\end{equation}
This corresponds to the sum of diagrams in Fig.~\ref{fiii}
with the quark propagators taken as  normal Feynman ones.

\section{Conclusions}
The main result of this paper is the explicit form of the
amplitude for the production of a real gluon off three
scattering centers, which is given by Eqs.
\Ref{e51}, \Ref{e52} and \Ref{e57}. An essential part of this
amplitude is the vertex R$\to$RRRP for the emission of a gluon
with the reggeon splitting into three calculated in Section 3.
The found amplitude can be  used for the calculation of
such observables as the inclusive cross-section for
gluon jet production on two nucleons in the nucleus
and the diffractive gluon production on three nucleons in the nucleus.

Two important properties have been established. First, we have
shown that in the sum of all terms principal
value singularities in the longitudinal momenta coming from the
induced contributions  combine with the
$\delta$-like singularities in the remaining contributions with multiple
interactions of the projectile to restore normal propagators for
the latter. So in the end no trace of the principal value prescription
remains. Second, we have found that in the kinematics appropriate for
the calculation of the  inclusive cross-section when the ''$-$''
component of the observed gluon momentum is much smaller than the
''$-$'' components of the momenta transferred to the target, the
vertex R$\to$RRRP is radically simplified and reduced to the
contribution of the diagram from Fig.~\ref{multi}.
Both results may be useful in the calculation of the inclusive
cross-sections.

Note that these properties are quite similar to the ones established
earlier for the production on two centers in \cite{BLSV}.
This makes us believe that they have a general validity
and are fulfilled for any number of scattering centers.

\section{Acknowledgments}
This work has been partially supported by the RFFI grant
12-02-00356-a .

\newpage
\section{Appendix A. Calculation of the induced vertex}
\renewcommand{\theequation}{\hbox{A.\arabic{equation}}}
\setcounter{equation}{0}

We denote the term
\begin{equation}
\frac{tr(T^{b}T^{b_1}T^{b_2}T^{b_3}T^{b_4})}
{(q_{2-}+q_{3-}+q_{4-})(q_{3-}+q_{4-})q_{4-}}
=-\frac{tr(T^{b}T^{b_1}T^{b_2}T^{b_3}T^{b_4})}
{q_{1-}(q_{3-}+q_{4-})q_{4-}}
\label{a3}
\end{equation}
as
$$
-\frac{[b1234]}{1(3+4)4}
$$
for brevity.

We have to calculate the convolution of the induced vertex
with the polarization vector of the gluon.
Then the common factor
\begin{equation}
-ig^3 (q^2_{\perp}) \epsilon^{*}_{-}=
ig^3 \frac{2(p\epsilon^{*})_{\perp}}{p_{+}} q^2_{\perp}
\label{a4}
\end{equation}
includes $(-1)$ from \Ref{a3}, $i$ for the vertex,
$-g^3$ from \Ref{a2}, $(-i)^5$ from fields, $-q^2_{\perp}$ from
$\partial^2_{\perp}$, $(-i)^3$ from three derivatives
$\partial^{-1}_{-}$ and the component
$\epsilon^{*}_{-}(p)=-2(p\epsilon^{*})_{\perp}/p_{+}\,$
of the polarization vector is correspondent to $V_{-}$.

Suppressing the factor \Ref{a4} and collecting similar terms,
for the vertex we get
$$
\frac{[b4321-b1234]}{1(1+2)4}
+\frac{[b3421-b1243]}{1(1+2)3}
+\frac{[b1423-b3241]}{1(2+3)3}
+\frac{[b1432-b2341]}{1(2+3)2}
$$
$$
+\frac{[b4231-b1324]}{1(1+3)3}
+\frac{[b2431-b1342]}{1(1+3)2}
+\frac{[b4312-b2134]}{2(1+2)4}
+\frac{[b3412-b2143]}{2(1+2)3}
$$
\begin{equation}
+\frac{[b4123-b3214]}{4(2+3)3}
+\frac{[b4132-b2314]}{4(2+3)2}
+\frac{[b4213-b3124]}{3(1+3)4}
+\frac{[b2413-b3142]}{3(1+3)2}\ .
\end{equation}
Further, using these identities:
$$
\frac{1}{(q_{1-}+q_{2-})q_{1-}}
=\frac{1}{q_{1-}q_{2-}}-\frac{1}{(q_{1-}+q_{2-})q_{2-}}\ ,
\ \ \frac{1}{(q_{1-}+q_{3-})q_{3-}}
=\frac{1}{q_{1-}q_{3-}}-\frac{1}{(q_{1-}+q_{3-})q_{1-}}\ ,
$$
$$
\frac{1}{(q_{2-}+q_{3-})q_{2-}}
=\frac{1}{q_{2-}q_{3-}}-\frac{1}{(q_{2-}+q_{3-})q_{3-}}\ ,
$$
we get
$$
\frac{[b4312-b2134-b4321+b1234]}{2(1+2)4}+
\frac{[b3412-b2143-b3421+b1243]}{2(1+2)3}
$$
$$
+\frac{[b4123-b3214-b4132+b2314]}{3(2+3)4}
+\frac{[b1423-b3241-b1432+b2341]}{1(2+3)3}
$$
$$
+\frac{[b4231-b1324-b4213+b3124]}{1(1+3)4}
+\frac{[b2431-b1342-b2413+b3142]}{1(1+3)2}
$$
$$
+\frac{[b4321-b1234]}{124}
+\frac{[b4132-b2314]}{234}
+\frac{[b4213-b3124]}{134}
$$
\begin{equation}
+\frac{[b3421-b1243+b1432-b2341+b2413-b3142]}{123}\ .
\label{a5}
\end{equation}

We can rewrite \Ref{a5} in the form of the expression
containing the factor $p_{-}=q_{4-}$ in the denominator
of each term with the use of the identities
$$
\frac{1}{(q_{1-}+q_{2-})q_{3-}}
=-\frac{1}{q_{3-}q_{4-}}-\frac{1}{(q_{1-}+q_{2-})q_{4-}}\ ,
\ \ \frac{1}{(q_{2-}+q_{3-})q_{1-}}
=-\frac{1}{q_{1-}q_{4-}}-\frac{1}{(q_{2-}+q_{3-})q_{4-}}\ ,
$$
$$
\frac{1}{(q_{1-}+q_{3-})q_{2-}}
=-\frac{1}{q_{2-}q_{4-}}-\frac{1}{(q_{1-}+q_{3-})q_{4-}}\ ,
$$
$$
\frac{1}{q_{1-}q_{2-}q_{3-}}=-\frac{1}{q_{1-}q_{2-}q_{4-}}
-\frac{1}{q_{1-}q_{3-}q_{4-}}-\frac{1}{q_{2-}q_{3-}q_{4-}}\ ,
$$
which are consequences of the relation
$q_{1-}+q_{2-}+q_{3-}+q_{4-}=0$.
Besides, we can note that the $T$-matrices
in the numerators form commutators $[T^a,T^b]=if^{abc}T^c$.
This gives
$$
if^{12c}\frac{[bc34+b43c-bc43-b34c]}{2(1+2)4}
+if^{23c}\frac{[bc14+b41c-bc41-b14c]}{3(2+3)4}
$$
$$
+if^{31c}\frac{[bc24+b42c-bc42-b24c]}{1(1+3)4}
$$
$$
-\frac{[b2431-b1342+b3421-b1243+b1432-b2341-b4321+b1234]}{124}
$$ $$
-\frac{[b3412-b2143+b1432-b2341+b2413-b3142-b4132+b2314]}{234}
$$
\begin{equation}
-\frac{[b1423-b3241+b3421-b1243+b2413-b3142-b4213+b3124]}{134}
\label{a6}
\end{equation}
for the vertex, where, for example, $f^{12c}\equiv f^{b_1b_2c}$.

Arranging the $T$-matrices in the commutators once and once more
and substituting \\
$tr([T^a,T^b]T^c )=\frac{i}{2}f^{abc}$, we get
$$
\frac{1}{q_{4-}}\left(
-f^{12c}f^{34d}\frac{[bcd-bdc]}{2(1+2)}
-f^{23c}f^{41d}\frac{[bdc-bcd]}{3(2+3)}
-f^{31c}f^{42d}\frac{[bdc-bcd]}{1(1+3)}
\right.
$$
$$
-if^{43d}\frac{[b2d1-b12d+b1d2-bd21]}{12}
-if^{14d}\frac{[-b2d3+b23d-b3d2+bd32]}{23}
$$ $$
\left.
-if^{42d}\frac{[b1d3-bd13+b3d1-b31d]}{13}
\right)=
$$
$$
=\frac{i}{2p_{-}}\left(
\frac{f^{12c}f^{43d}f^{bcd}}{2(1+2)}+
\frac{f^{23c}f^{41d}f^{bcd}}{3(2+3)}+
\frac{f^{31c}f^{42d}f^{bcd}}{1(1+3)}+
\right.
$$
\begin{equation}
\left.
+\frac{f^{2cd}f^{ca1}f^{43d}}{12}
+\frac{f^{3cd}f^{ca2}f^{41d}}{23}
+\frac{f^{1cd}f^{ca3}f^{42d}}{13}
\right)\ .
\label{a7}
\end{equation}
Restoring the original notation and taking into account the factor
\Ref{a4}, we finally get the answer
$$
V_5^{ind}=
g^3 \frac{(p\epsilon^{*})_{\perp}}{p^2_{\perp}} q^2_{\perp}
\left(
\frac{f^{bcd}f^{b_1b_2c}f^{ab_3d}}
{(q_{1-}+q_{2-})q_{2-}}
+\frac{f^{bcd}f^{b_2b_3c}f^{ab_1d}}
{(q_{2-}+q_{3-})q_{3-}}
+\frac{f^{bcd}f^{b_3b_1c}f^{ab_2d}}
{(q_{1-}+q_{3-})q_{1-}} +\right.
$$
\begin{equation}
\left. +\frac{f^{bb_1c}f^{b_2cd}f^{ab_3d}}
{q_{1-}q_{2-}}
+\frac{f^{bb_2c}f^{b_3cd}f^{ab_1d}}
{q_{2-}q_{3-}}
+\frac{f^{bb_3c}f^{b_1cd}f^{ab_2d}}
{q_{1-}q_{3-}} \right) \, .
\label{a8}
\end{equation}
On the mass shell of the induced gluon the factor
$p_{+}p_{-}=-p^2_{\perp}$ appears in the denominator,
therefore the vertex remains finite in the limit $p_{-}\to 0$.

The vertex can be expressed in the more convenient form. With the use
of the Jacobi identity
$
f^{bcd}f^{b_1b_2c}=f^{bb_2c}f^{b_1cd}-f^{bb_1c}f^{b_2cd}
$
and the relation
$$
\frac{1}{q_{1-}q_{2-}}=
\frac{1}{(q_{1-}+q_{2-})q_{1-}}+\frac{1}{(q_{1-}+q_{2-})q_{2-}}
$$
the sum of the first and the fourth term in \Ref{a8}
can be rewritten as the sum
$$
\frac{f^{bb_1c}f^{b_2cd}f^{ab_3d}}{(q_{1-}+q_{2-})q_{1-}}
+\frac{f^{bb_2c}f^{b_1cd}f^{ab_3d}}{(q_{1-}+q_{2-})q_{2-}}\ .
$$
With an analogous transformation of the rest terms,
the vertex takes the form
$$
V_5^{ind}=
g^3 \frac{(p\epsilon^{*})_{\perp}}{p^2_{\perp}} q^2_{\perp}
\left(
\frac{f^{bb_1c}f^{cb_2d}f^{db_3a}}{q_{1-}(q_{1-}+q_{2-})}
+\frac{f^{bb_2c}f^{cb_3d}f^{db_1a}}{q_{2-}(q_{2-}+q_{3-})}
+\frac{f^{bb_3c}f^{cb_1d}f^{db_2a}}{q_{3-}(q_{1-}+q_{3-})}
\right.
$$
\begin{equation}
\left.
+\frac{f^{bb_3c}f^{cb_2d}f^{db_1a}}{q_{3-}(q_{2-}+q_{3-})}
+\frac{f^{bb_2c}f^{cb_1d}f^{db_3a}}{q_{2-}(q_{1-}+q_{2-})}
+\frac{f^{bb_1c}f^{cb_3d}f^{db_2a}}{q_{1-}(q_{1-}+q_{3-})}
\right) \, .
\label{afin}
\end{equation}

The expression \Ref{afin} is obviously symmetrical with respect to
all permutations of momenta $q_{1,2,3}$ and correspondent indices
$b_{1,2,3}$ of the induced reggeons.

\section{Appendix B. Calculation of the four-reggeon vertex}
\renewcommand{\theequation}{\hbox{B.\arabic{equation}}}
\setcounter{equation}{0}

Since the quadruple term $tr([v_{+},v_{-}][v_{-},v_{-}])$ is absent
in the Yang-Mills Lagrangian, the four-reggeon vertex
$<A_{+}A_{-}A_{-}A_{-}>$ arises only from the term
\begin{equation}
g^2 tr(A_{+}\partial^2_{\perp}
v_{-}\partial^{-1}_{-}v_{-}\partial^{-1}_{-}v_{-})
=g^2 tr(A_{+}\partial^2_{\perp}
A_{-}\partial^{-1}_{-}A_{-}\partial^{-1}_{-}A_{-}) +\dots
\label{e71}
\end{equation}
of the induced part of the Lagrangian density \Ref{ei1}.

To write the vertex in the momentum representation
\begin{equation}
ig^2
\frac{tr(T^{c}T^{b_3}T^{b_2}T^{b_1})} {q_{1-}(q_{1-}+q_{2-})}
\cdot (q_1 +q_2 +q_3)^2_{\perp}
+ \Big(\mbox{permutations of } q_{1,2,3} \mbox{ and } b_{1,2,3}\Big) .
\label{e72}
\end{equation}
we took into account the following factors: $i$ for the vertex,
$g^2$ from \Ref{e71}, $(-i)^4$ from fields,
$-(q_1 +q_2 +q_3)^2_{\perp}$ from $\partial^2_{\perp}$ and
$(-i)^2$ from two derivatives $\partial^{-1}_{-}$.

Suppressing the common factor and denoting the typical term as
$$
\frac{tr(T^{c}T^{b_3}T^{b_2}T^{b_1})} {q_{1-}(q_{1-}+q_{2-})}
\equiv \frac{[c321]}{1(1+2)}
$$
we rewrite \Ref{e72} in the form
\begin{equation}
\frac{[c321]}{1(1+2)}
+\frac{[c213]}{3(1+3)}
+\frac{[c132]}{2(2+3)}
+\frac{[c123]}{3(2+3)}
+\frac{[c231]}{1(1+3)}
+\frac{[c312]}{2(1+2)} \ .
\label{e73}
\end{equation}
Using the relation $q_{1-}+q_{2-}+q_{3-}=0$
and the identity
$$
\frac{1}{q_{1-}q_{2-}}=
\frac{1}{q_{1-}(q_{1-}+q_{2-})}+\frac{1}{q_{2-}(q_{1-}+q_{2-})}
$$
we get
$$
\frac{[c321+c123-c132-c231]}{1(1+2)}
+\frac{[c312+c213-c132-c231]}{2(1+2)}
$$
$$
=
\frac{i f^{db_3 b_2} [cd1-c1d]} {1(1+2)}
+\frac{i f^{db_1 b_3} [c2d-cd2]} {2(1+2)}
$$
\begin{equation}
=\frac{1}{2}
\left(
\frac{f^{cb_1 d}f^{db_3 b_2}}{q_{1-}(q_{1-}+q_{2-})}
+\frac{f^{cb_2 d}f^{db_3 b_1}}{q_{2-}(q_{1-}+q_{2-})}
\right) \ .
\label{e74}
\end{equation}

In the original notation, restoring the common factor,
we get finally for the vertex
\begin{equation}
\frac{ig^2}{2}
\left(
\frac{f^{cb_1 d}f^{db_3 b_2}}{q_{1-}(q_{1-}+q_{2-})}
+\frac{f^{cb_2 d}f^{db_3 b_1}}{q_{2-}(q_{1-}+q_{2-})}
\right) \cdot (q_1 +q_2 +q_3)^2_{\perp} \ .
\label{e7a}
\end{equation}

The expression for the diagram from Fig.~\ref{multi} is given
by the four-reggeon vertex, multiplied by the Lipatov vertex,
two reggeon propagators and the quark-reggeon vertex and equals
\begin{equation}
g^4 \gamma_{+} L(p,q_{1}+q_{2}+q_{3}) T^{b}
\left(
\frac{f^{bac}f^{cb_1 d}f^{db_3 b_2}}{q_{1-}(q_{1-}+q_{2-})}
+\frac{f^{bac}f^{cb_2 d}f^{db_3 b_1}}{q_{2-}(q_{1-}+q_{2-})}
\right) \ ,
\label{e70}
\end{equation}
what coincides with \Ref{e6a}.

\section{Appendix C. Proof of transversality of the R$\to$RRRP vertex}
\renewcommand{\theequation}{\hbox{C.\arabic{equation}}}
\setcounter{equation}{0}

\subsection{Effective vertex in an arbitrary gauge}
To verify the transversality of R$\to$RRRP vertex it's necessary to
write for it the explicit expression in an arbitrary gauge. As it was shown
such an expression consists of several parts.
The part corresponding to fig.~\ref{fi}.1 is
\begin{equation}
\begin{split}
\frac{g f^{b_3 ad}}{2}
\Big( -2p_+ g_{\mu\sigma} +(p+2q_3)_{\mu} n^+_{\sigma}
+(p-q_3)_{\sigma} n^+_{\mu} +\frac{q_3^2}{p_+}n^+_{\mu}n^+_{\sigma}
\Big)\\
\times \frac{-i}{(q-q_{1}-q_{2})^2 +i0} \cdot
\frac{g f^{b_2 dc}}{2}
\Big( -2r_+ g_{\sigma\lambda} +(r+2q_2)_{\sigma} n^+_{\lambda}
+(r-q_2)_{\lambda} n^+_{\sigma} +\frac{q_2^2}{r_+}n^+_{\sigma}n^+_{\lambda}
\Big)\\
\times \frac{-i}{(q-q_{1})^2 +i0} \cdot \frac{g f^{bb_1 c}}{2}
\left[ q_{\lambda}+q_{1\lambda}
+\left( \frac{q_1^2}{q_+} -q_{1-} \right) n^+_{\lambda}
+\left( \frac{q^2}{q_{1-}} -q_+ \right) n^-_{\lambda} \right]=
\\
-\frac{g^3f^{b_3 ad}f^{b_2 dc}f^{bb_1 c}}{8[(q-q_{1})^2 +i0][(q-q_1-q_2)^2+i0]}
(4p_+^2g_{\mu\lambda}+p_+n_\lambda^+(-2r-4q_2-p-2q_3)_\mu+\\
p_+n_\mu^+(-3r+q_2+4q_3)_\lambda+n_\mu^+n_\lambda^+(-q_2^2-2q_3^2+p^2+2q_2p-2q_2q_3))\\
\times \frac{-i}{(q-q_{1})^2 +i0} \cdot \frac{g f^{bb_1 c}}{2}
\left[ q_{\lambda}+q_{1\lambda}
+\left( \frac{q_1^2}{q_+} -q_{1-} \right) n^+_{\lambda}
+\left( \frac{q^2}{q_{1-}} -q_+ \right) n^-_{\lambda} \right]=
\\
-\frac{g^3f^{b_3 ad}f^{b_2 dc}f^{bb_1 c}}{8[(q-q_{1})^2 +i0][(q-q_1-q_2)^2+i0]}
(4p_+^2(-2r-4q_2-p-2q_3)_\mu+p_+n_\mu^+(q_2+q_3-3p)(q+q_1)+\\
p_+n_\mu^+(-q_2^2-2q_3^2+p^2+2q_2p-2q_2q_3)+p_+^2n_\mu^
 +\left( \frac{q_1^2}{q_+} -q_{1-} \right)+\\
\left( \frac{q^2}{q_{1-}} -q_+ \right)(2p_+(-2r-4q_2-p-2q_3)_\mu+4p_+^2n_\mu^-+\\
p_+n_\mu^+(-3r+q_2+4q_3)_-+2n_\mu^+(-q_2^2-2q_3^2+p^2+2q_2p-2q_2q_3))) .
\label{cc1}
\end{split}
\end{equation}
The part corresponding to fig.~\ref{fi}.2 and fig.~\ref{fi}.3 is
\begin{equation}
\begin{split}
\frac{g f^{b_3 ad}}{2}
\Big( -2p_+ g_{\mu\sigma} +(p+2q_3)_{\mu} n^+_{\sigma}
+(p-q_3)_{\sigma} n^+_{\mu} +\frac{q_3^2}{p_+}n^+_{\mu}n^+_{\sigma}
\Big)
\\
\times
\frac{-i}{(q-q_{1}-q_{2})^2 +i0}
\frac{ig^2}{4}
f^{bb_1 c}f^{cb_2 d}\left[ -\frac{2q_\perp^2 n_\sigma^-}
{(q_1 +q_2)_{-}q_{1-}} -n_\sigma^+ \right]=
\\
\frac{g^3f^{b_3 ad}f^{b_2 dc}f^{bb_1 c}}{8[(q-q_1-q_2)^2+i0]}
\!\!\left(\!\!
\frac{4p_+q_\bot^2n_\mu^-}{(q_1+q_2)_-q_{1-}}
-\frac{4q_\bot^2(p+2q_3)_\mu}{(q_1+q_2)_-q_{1-}}
-\frac{2(p-q_3)_-q_\bot^2n_\mu^+}{(q_1+q_2)_-q_{1-}}
-\frac{4q_3^2q_\bot^2n_\mu^+}{p_+(q_1+q_2)_-q_{1-}}+p_+n_\mu^+
\!\!\right)\!\!.
\end{split}
\label{cc2}
\end{equation}
The part corresponding to fig.~\ref{fi}.4 is
\begin{equation}
\begin{split}
\frac{-ig^2}{4}
 f^{cb_2 d}f^{db_3 a}
n^{+}_{\mu}n^{+}_{\lambda}\\
\times \frac{-i}{(q-q_{1})^2 +i0} \cdot \frac{g f^{bb_1 c}}{2}
\left[ q_{\lambda}+q_{1\lambda}
+\left( \frac{q_1^2}{q_+} -q_{1-} \right) n^+_{\lambda}
+\left( \frac{q^2}{q_{1-}} -q_+ \right) n^-_{\lambda} \right]=\\
\frac{g^3f^{b_3 ad}f^{b_2 dc}f^{bb_1 c}}{8[(q-q_1)^2+i0]}
\left(p_+n_\mu^+-\frac{2q_\bot^2n_\mu^+}{q_{1-}}
\right)\, .
\end{split}
\label{cc3}
\end{equation}
The part corresponding to fig.~\ref{fi}.5 is
\begin{equation}
\begin{split}
-ig^3q_{\bot}^2n_{\mu}^-
\frac{i}{2p_-}
\frac{f^{b_3 ad}f^{b_2dc}f^{bb_1c}}{q_{1-}(q_{2-}+q_{1-})}\ .
\label{cc4}
\end{split}
\end{equation}
So the total expression for $V_5$-effective vertex in an arbitrary
gauge is equal to the sum of expressions \Ref{cc1}-\Ref{cc4}.

\subsection{Proof of transversality}
To prove the transversality of effective $V_5$-vertex one should convolute
an expression for it in arbitrary gauge with
momentum vector $p_\mu$. We will do it separately for each of the parts
\Ref{cc1}-\Ref{cc4}. Part \Ref{cc1} convoluted with $p_\mu$
gives:
\begin{equation}
\begin{split}
\frac{g^3f^{b_3 ad}f^{b_2 dc}f^{bb_1 c}}{8[(q-q_{1})^2 +i0][(q-q_1-q_2)^2+i0]}
\left(
\frac{2q_\bot^2p_+}{q_{1-}}-p_+^2
\right)
\left(
(q-q_{1})^2+(q-q_1-q_2)^2
\right)\, .
\label{cc5}
\end{split}
\end{equation}
For \Ref{cc2}-\Ref{cc3} we correspondingly have:
\begin{equation}
\begin{split}
\frac{g^3f^{b_3 ad}f^{b_2 dc}f^{bb_1 c}}{8[(q-q_{1})^2 +i0][(q-q_1-q_2)^2+i0]}
\left(
\frac{2q_\bot^2p_+}{q_{1-}}+p_+^2
\right)
(q-q_1)^2 \ ,
\end{split}
\label{cc6}
\end{equation}
\begin{equation}
\begin{split}
\frac{g^3f^{b_3 ad}f^{b_2 dc}f^{bb_1 c}}{8[(q-q_{1})^2 +i0][(q-q_1-q_2)^2+i0]}
\left(
\frac{-2q_\bot^2p_+}{q_{1-}}+p_+^2
\right)
(q-q_1-q_2)^2 \ .
\end{split}
\label{cc7}
\end{equation}
For \Ref{cc4} we have:
\begin{equation}
\begin{split}
-\frac{g^3f^{b_3 ad}f^{b_2 dc}f^{bb_1 c}}{8[(q-q_{1})^2 +i0][(q-q_1-q_2)^2+i0]}
\frac{4q_\bot^2p_+}{q_{1-}}
(q-q_1)^2 \ .
\label{cc8}
\end{split}
\end{equation}
So the sum \Ref{cc5}-\Ref{cc8} is zero.

\end{document}